\documentclass[twocolumn]{aastex62}
\usepackage{amsmath,amssymb}
\usepackage{multirow}
\hypersetup{linkcolor=red,citecolor=green,filecolor=cyan,urlcolor=magenta}

\received{xxx, xxxx}
\revised{xxx, xxxx}
\accepted{xxx, xxxx}
\submitjournal{ApJ}

\shorttitle{}
\shortauthors{S. Zha et al.}

\begin{document}

\title{Accretion-Induced Collapse of Dark Matter Admixed White Dwarfs - II: Rotation and Gravitational-wave Signals}

\correspondingauthor{Shuai Zha}
\email{szha@phy.cuhk.edu.hk}

\author{Shuai Zha}
\author{Ming-chung Chu}
\affiliation{Department of Physics and Institute of Theoretical Physics, The Chinese University of Hong Kong, \\Shatin, N.T., Hong Kong S.A.R., China}

\author{Shing-chi Leung}
\affiliation{Kavli Institute for the Physics and Mathematics of the Universe
	(WPI), The University of Tokyo Institutes for Advanced Study, \\The University of Tokyo, Kashiwa, Chiba 277-8583, Japan}

\author{Lap-ming Lin}
\affiliation{Department of Physics and Institute of Theoretical Physics, The Chinese University of Hong Kong, \\Shatin, N.T., Hong Kong S.A.R., China}



\begin{abstract}

We present axisymmetric hydrodynamical simulations of accretion-induced collapse (AIC) of dark matter (DM) admixed rotating white dwarfs (WD) and their burst gravitational-wave (GW) signals. For initial WD models with the same central baryon density, the admixed DM is found to delay the plunge and bounce phases of AIC, and decrease the central density and mass of the proto-neutron star (PNS) produced. The bounce time, central density and PNS mass generally depend on two parameters, the admixed DM mass $M_\mathrm{DM}$ and the ratio between the rotational kinetic and gravitational energies of the inner core at bounce $\beta_\mathrm{ic,b}$. The emitted GWs have generic waveform shapes and the variation of their amplitudes $h_+$ show a degeneracy on $\beta_\mathrm{ic,b}$ and $M_\mathrm{DM}$. We found that the ratios between the GW amplitude peaks around bounce allow breaking the degeneracy and extraction of both $\beta_\mathrm{ic,b}$ and $M_\mathrm{DM}$. Even within the uncertainties of nuclear matter equation of state, a DM core can be inferred if its mass is greater than 0.03 $M_{\odot}$. We also discuss possible DM effects on the GW signals emitted by PNS g-mode oscillations. GW may boost the possibility for the detection of AIC, as well as open a new window in the indirect detection of DM. 

\end{abstract}



\section{Introduction} \label{sec:intro}
It is commonly known that dark matter (DM) constitutes approximately 84\% of the matter in the universe \citep{2018arXiv180706209P}. The existence of DM is crucial for explaining the flattening of galactic rotation curve \citep{1974Natur.252..111E} and the Bullet cluster observation \citep{2010ApJ...718...60L}, but the nature of DM is largely unknown despite decades of searches \citep[e.g., ][]{2018PhRvL.121b1304R} and the many hypothetical candidates \citep{2018RvMP...90d5002B}. DM is believed to play an important role in the formation of cosmic microwave background anisotropies (CMBA) and large-scale structures. An interesting question is what impacts they might have on small-scale structures, such as stars and supernovae. DM admixed in star provide extra gravity to alter the stellar structure \citep{2015PhRvL.115k1301B} and additional heating/cooling to alter its surface luminosity and lifespan \citep{2015PhRvL.115n1301B,2017A&A...605A.106C}.

With a very high central density ($\rho_\mathrm{c}$), compact stars such as white dwarfs (WD, $\rho_\mathrm{c}\sim 10^{9-10}~\mathrm{g/cm^3}$) and neutron stars (NS, $\rho_\mathrm{c}\sim10^{14-15}~\mathrm{g/cm^3}$) provide great observational playgrounds for detecting DM \citep{2011PhRvD..84j7301L,2013PhRvD..87l3506L,2018PhRvD..98k5027G,2018PhRvD..97l3007E}, complementary to direct detection experiments. \cite{2015MNRAS.450L..71F} proposed that DM-induced collapse of NSs can explain the missing pulsar problem at the galactic center and the fast radio burst phenomena. Also, there have been studies on the effects of DM on the thermonuclear explosions of WDs, i.e., Type Ia supernovae (SNe~Ia). \cite{2015PhRvL.115n1301B} studied the accretion of PeV-scale DM by WDs and found that this can explain the age-luminosity anti-correlation relation for SNe~Ia. \cite{2015PhRvD..92f3007G} suggested that the transit of primordial black holes (PBH) can ignite SNe~Ia through heating by dynamical friction and the scenario can put some constraints on the PBH mass. \cite{2015ApJ...812..110L} found that DM admixture decreases the canonical WD Chandrasekhar mass ($M_\mathrm{Ch}\sim1.44~M_\odot$), and their subsequent hydrodynamical simulations showed that DM admixed SNe Ia synthesize less $^{56}\mathrm{Ni}$ and can fit some sub-luminous SN~Ia light curves.

While SNe~Ia have been studied extensively \citep[see the review in ][]{2013FrPhy...8..116H}, accretion-induced collapse (AIC) has been proposed as an alternative fate of WDs approaching $M_\mathrm{Ch}$ \citep{1991ApJ...367L..19N}. Massive oxygen-neon-magnesium (ONeMg) WDs left behind by intermediate-mass stars ($8-9~M_\odot$) are thought to follow this pathway more probably \citep{2017MNRAS.472.3390S}. Though not confirmed in electromagnetic observations yet, AIC has several important theoretical and observational implications. Compared to core-collapse supernovae (CCSNe) of massive stars ($M\gtrsim10~M_\odot$), AICs leave behind remnant NSs with a lower baryonic mass ($\sim 1.35~M_\odot$) as the progenitor WDs weigh $M_\mathrm{Ch}\sim1.44~M_\odot$ at most, ignoring rotational effects. These NSs can be the low-mass branch pulsars found in the bimodal NS mass distribution \citep{2010ApJ...719..722S,2012ApJ...757...55O,2019ApJ...876...18F}. AICs have interesting nucleosynthesis patterns and may contribute to the production of silver, palladium \citep{2012A&A...545A..31H}, and some r-process elements \citep{1999ApJ...516..892F,2019A&A...622A..74J}. Due to the thin envelop and weak explosion energy ($\lesssim 10^{50}~\mathrm{erg}$), AICs exhibit themselves as short and faint transients in electromagnetic waves, while they can emit strong X-ray flashes lasting $\sim 1~\mathrm{hr}$ in binary systems \citep{2014ApJ...794...28P}. A natural question is how DM admixture will change the evolution and outcomes of AIC.

To approach $M_\mathrm{ch}$, the progenitor WD accretes mass and angular momentum from its companion, and so rotation is an important ingredient of SNe~Ia and AICs. \cite{2018A&A...618A.124F} pointed out that thermonuclear explosions of rapidly rotating WDs can be candidates for superluminous SNe~Ia. The collapses of  rotating stars, including AICs, are expected to emit strong bursts of gravitational waves (GW) \citep{2009CQGra..26f3001O}. They are among the potential targets for ground-based GW detectors such as LIGO\ \footnote{\url{www.ligo.org/}}, Virgo, \footnote{\url{public.virgo-gw.eu/}}, KAGRA\ \footnote{\url{gwcenter.icrr.u-tokyo.ac.jp}}, and the third generation detector Einstein Telescope \citep{2011CQGra..28i4013H}. Due to the complicated nature of the collapse and bounce dynamics, there is no analytic solution for the GW waveforms for rotating AICs, which are indispensable for powerful data analysis algorithms such as matched filtering \citep{2016PhRvD..93d2002G}. Accurate waveforms can only be calculated from computationally demanding hydrodynamical simulations. Because AICs emit transients of optical photons \citep{2014ApJ...794...28P}, radio-waves \citep{2013ApJ...762L..17P}, neutrinos \citep{2006ApJ...644.1063D} and GWs, they are  interesting candidates in the new era of multi-messenger astronomy. An extended question is how these observations can tell us whether DM is admixed in the progenitor or not.

In paper I \citep{2019LeungAIC}, we performed spherically symmetric simulations and found that DM admixed AICs produce light NSs with mass compatible with that of the recently observed low mass (baryonic mass $\sim1.28~M_\odot$) NS in the binary system PSR J0453+1559 \citep{2015ApJ...812..143M}.  In this paper, we examine the AIC dynamics and GW signals when the rotating progenitor WD bears a DM core of different masses, with axisymmetric hydrodynamical simulations. Specifically, we check whether the presence of DM can be identified through the GW observations.

This paper is organized as follows. Section~\ref{sec:method} introduces the methods for constructing initial models, the subsequent hydrodynamical simulations and extraction of GW waveforms. Section~\ref{sec:results} presents the results of AIC simulations, focusing on the dependence of the dynamics and emitted GWs on both rotation rate and $M_\mathrm{DM}$. We discuss how to break the degeneracy of GWs on rotation rate and $M_\mathrm{DM}$ and retrieve parameters with a quantitative analysis of the extracted GW waveforms. In Section~\ref{sec:discuss}, we compare the difference between this work and a previous study without DM admixture \citep{2010PhRvD..81d4012A}. We also discuss the possible observational implications on the proto-neutron star (PNS) g-mode GW emission. We summarize our results in Section~\ref{sec:conclu}.
  
\section{Methods} \label{sec:method}
We performed axisymmetric simulations of AIC starting from a rotating WD with various $M_\mathrm{DM}$, using the Newtonian hydrodynamical code developed in \cite{2015MNRAS.454.1238L}. The code has been used to model SNe~Ia \citep{2015ApJ...812..110L,2018ApJ...861..143L,2019arXiv190110007L} and electron capture supernovae \citep{2019arXiv190111438L}, and we implemented necessary physics modules for modeling AIC and extracting GW waveforms. This section outlines the essence of the methods used in this paper. 

\subsection{Initial models} \label{subsec:init}
We followed \cite{1986ApJS...61..479H} to generate the rotating WD self-consistently as the initial model for the hydrodynamical simulations.  In this method, the equation of rotational equilibrium is given by
\begin{equation}
H(\rho) \equiv \int \rho^{-1}\mathrm{d}P = C-\Phi+\int \Omega^2 \varpi \mathrm{d}\varpi.  \label{eq:rotWD}
\end{equation}
Here, $\rho$ and $P$ are the density and pressure of baryonic normal matter (NM), $C$ the constant of integration, $\Phi$ the gravitational potential, $\varpi$ the perpendicular distance from the axis of rotation and $\Omega$ the angular velocity at $\varpi$. This integral equation is solved iteratively until the constant $C$ and stellar mass $M$ converge. The equation of state (EOS) of NM ($P=P(\rho)$) is the ideal degenerate electron gas EOS with an electron fraction $Y_e=0.5$. The extra gravity provided by the DM is described in the Poisson equation by including the DM density ($\rho_{\rm DM}$)
\begin{equation}
\nabla^2 \Phi = 4\pi G (\rho+\rho_\mathrm{DM}). \label{eq:poisson}
\end{equation}
The DM admixture is assumed to be non-rotating for simplicity and its structure together with non-rotating NM is initially solved by a spherically-symmetric two-fluid hydrostatic equation \citep{2015ApJ...812..110L}
\added{
\begin{equation}
\begin{aligned}
\frac{\mathrm{d}P}{\mathrm{d}r} &= - \frac{G(M(r)+M_\mathrm{DM}(r))}{r^2} \rho, \\
\frac{\mathrm{d}P_\mathrm{DM}}{\mathrm{d}r} &= - \frac{G(M(r)+M_\mathrm{DM}(r))}{r^2} \rho_\mathrm{DM},
\end{aligned}
\end{equation}
where $P_\mathrm{DM}$ is the pressure of DM calculated according to the DM model described below. $M(r)$ and $M_\mathrm{DM}(r)$ are the enclosed masses of NM and DM, respectively, at radius $r$ calculated by
\begin{equation}
\begin{aligned}
\frac{\mathrm{d}M(r)}{\mathrm{d}r} &= {4\pi r^2} \rho, \\
\frac{\mathrm{d}M_\mathrm{DM}(r)}{\mathrm{d}r} &= {4\pi r^2} \rho_\mathrm{DM}.
\end{aligned}
\end{equation}
} The DM density profile $\rho_\mathrm{DM}$ is then fixed during the iteration process of solving Eqs.~\ref{eq:rotWD} and \ref{eq:poisson}.~\footnote{We found that the change in  $\rho_\mathrm{NM}$ within the size of DM admixture is $\mathcal{O}(10^{-3})$.} 

Since the properties of DM are very uncertain \citep{2018RvMP...90d5002B}, as the first example in studying its effects on AIC, DM particles are assumed to be ideal Fermions with particle mass $1~\mathrm{GeV}$ without any self-annihilation or self-interaction \citep{2014PhRvC..89b5803X,2015PhRvD..92f3526K,2015ApJ...812..110L,2016PhRvD..93h3009M,2017PASA...34...43C}. The choice of particle mass $1~\mathrm{GeV}$ is consistent with a recent proposal that the mass of DM particle is less than a few GeV \citep{2018Natur.555...71B} in order to explain the amplitude of the 21-centimeter signal from hydrogen atoms in the very early universe detected by EDGES \citep{2018Natur.555...67B}. 

The rotation law for AIC or SN~Ia progenitors is uncertain and depends on how the WD grows to $M_\mathrm{Ch}$ \citep{2004A&A...419..623Y,2010PhRvD..81d4012A}. Also, the central density for an AIC progenitor is not accurately determined from stellar evolution calculation yet \citep{2017MNRAS.472.3390S, 2019ApJ...872..131S}, but it must lie in a narrow range for the gravitational instability to trigger the collapse. As we are focusing on the effects of DM admixture on AIC, as a first step only uniformly rotating progenitors (the angular velocity $\Omega$ is constant throughout the WD) are considered with a fixed central density of $5\times10^{10}~\mathrm{g/cm}^3$, which is a value commonly used in hydrodynamical simulations of AIC \citep{2006ApJ...644.1063D, 2010PhRvD..81d4012A} and close to the value $\sim4\times10^{10}~\mathrm{g/cm}^3$ for ONeMg core collapse \citep{2006A&A...450..345K}. We have used other central densities to verify that its exact value does not affect our conclusion on the DM effects in Appendix~\ref{app:rho_ci}, and we leave differential rotation for a future study. 

\added{
\begin{figure}
	\plotone{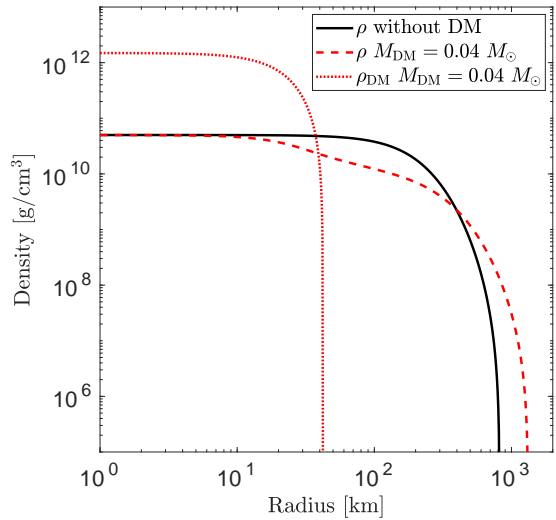}
	\caption{\added{Density profiles of NM and DM for two non-rotating WD models. The black solid line is the NM density profile for the WD without DM admixture. The red dashed and dotted lines are the NM and DM density profiles, respectively, for the WD with $0.04~M_\odot$ of DM admixture.} \label{fig:rho_nonrot}}
\end{figure}

Firstly, in Fig.~\ref{fig:rho_nonrot} we plot the density profiles of NM and DM for two non-rotating WD models, without DM and with DM admixture of $0.04~M_\odot$. For the DM admixed model, DM resides in the central region as a compact core, with a radius of $\sim40$ km.  The lower limit for integrating $M_{\rm DM}$ is taken to be $10^4 {\rm g/cm}^3$. The exact value of this lower limit does not significantly affect $M_{\rm DM}$ as the DM density drops steeply at $\rho_{\rm DM}\lesssim10^9$ g/cm$^3$. As the DM core provides additional gravitational pull to NM, the NM density decreases more quickly compared to the model without DM admixture. However, the WD becomes more extended due to the smaller total mass of the hybrid star, including NM and DM.
}

\begin{figure*}[t!]
	\plottwo{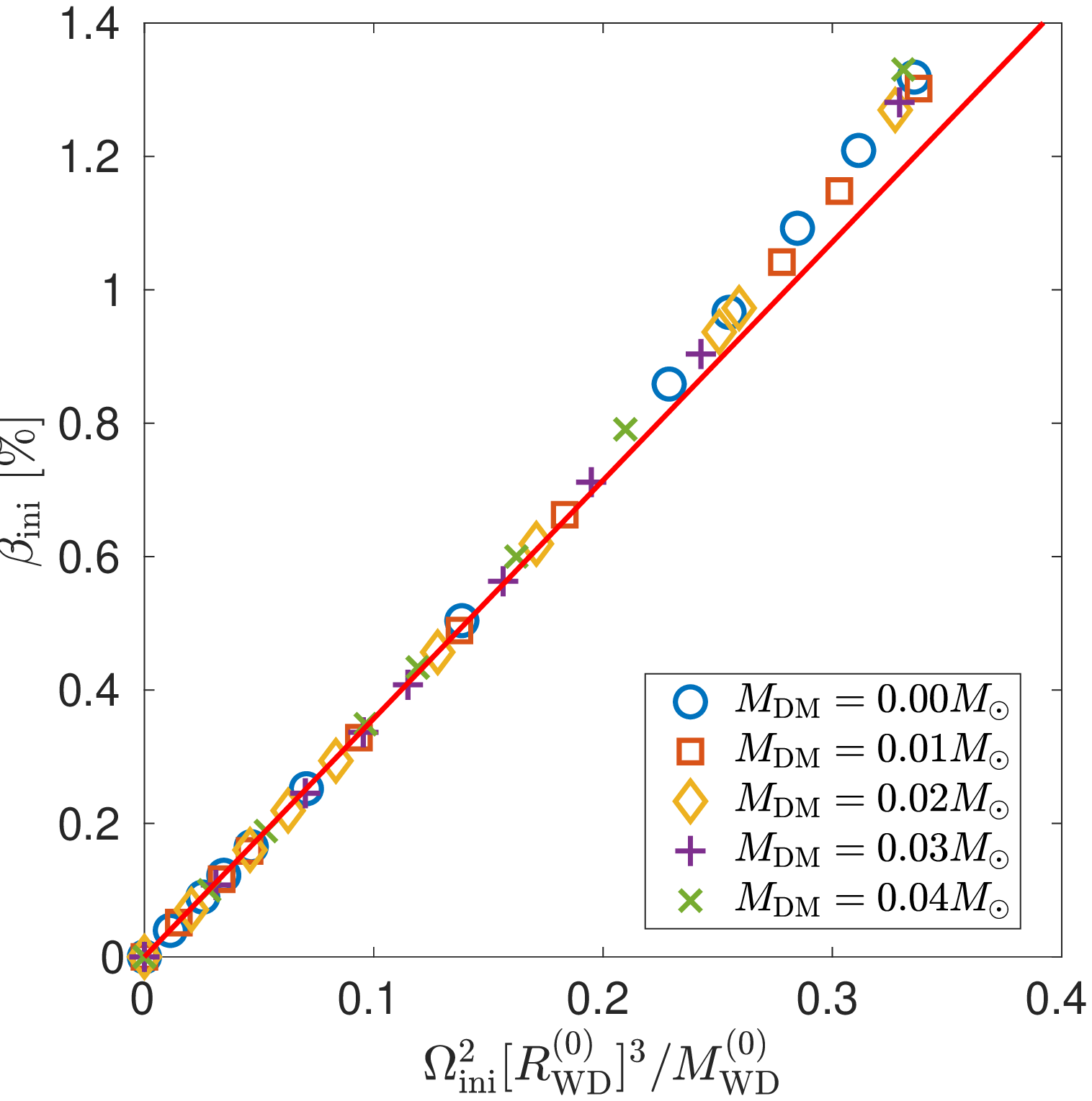}{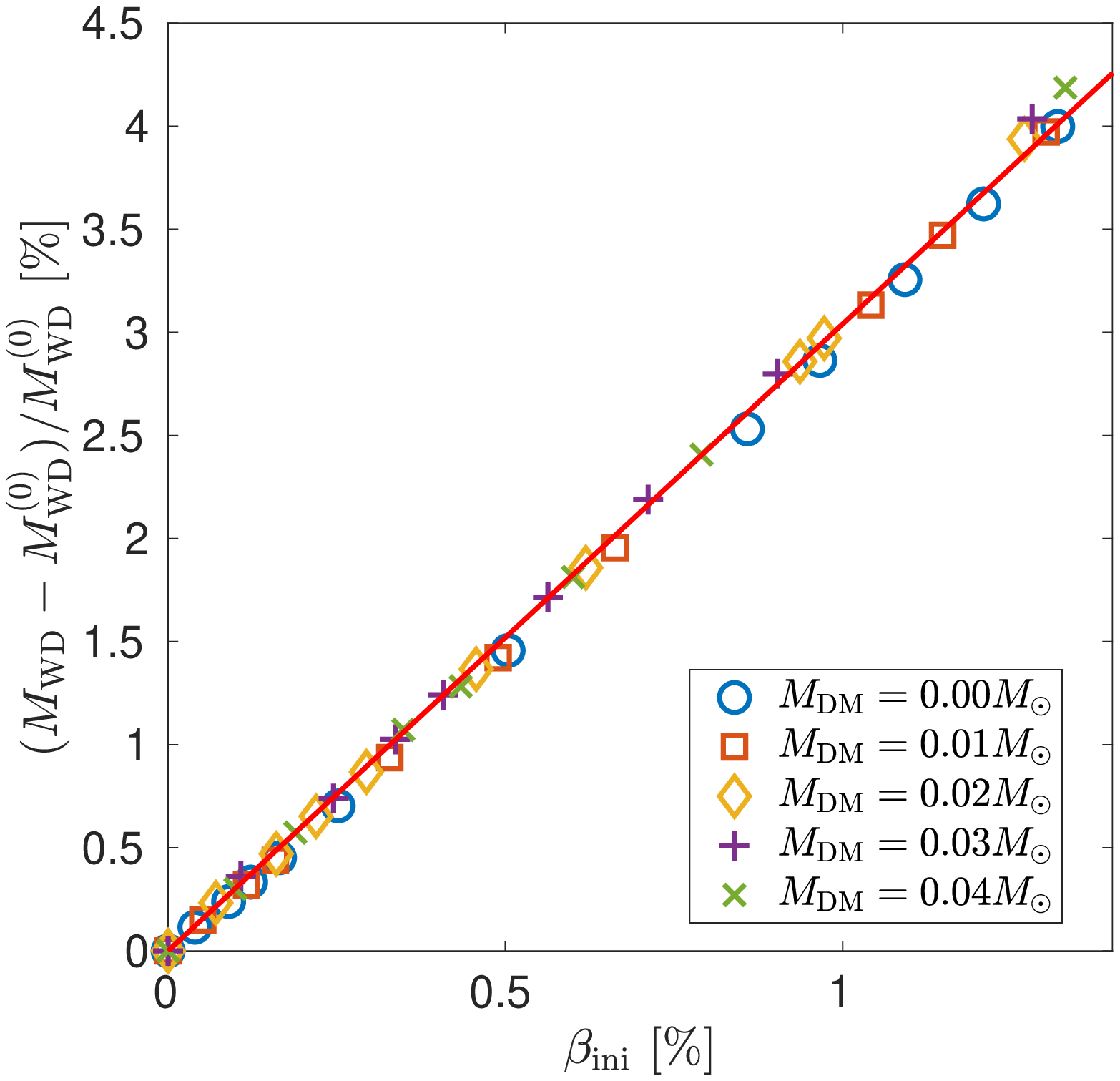}
	\caption{$\beta_\mathrm{ini}$ (left) and fractional increment of $M_\mathrm{WD}$ relative to $M_{\mathrm{WD}}^{(0)}$ in the non-rotating case (right) for the initial rotating WDs, with different $M_\mathrm{DM}$. $\beta_\mathrm{ini}$ shows a quadratic relation with $\Omega_\mathrm{ini}$ for slow rotation ($\beta_\mathrm{ini}\lesssim0.7\%$) and is proportional to the relative increment of $M_\mathrm{WD}$. \label{fig:init}}
\end{figure*}

The global properties of $1~\mathrm{GeV}$ fermionic DM admixed rotating WDs with the same initial angular velocity ($\Omega_\mathrm{ini}=5~\mathrm{rad/s}$, near the Keplerian limit for $M_\mathrm{DM}=0.04~M_\odot$) are summarized in Table \ref{tab:init_model_R5}. Admixture of DM makes the star more extended in radial size but less massive, and increases the ratio between the rotational kinetic and gravitational binding energies ($\beta\equiv E_\mathrm{Rot}/|E_\mathrm{Grav}|$). The total masses, $M_\mathrm{WD}$ (including NM and DM),  and initial values of the $\beta$ parameter, $\beta_\mathrm{ini}$, for all the constructed WDs are plotted in Fig.~\ref{fig:init}. For slow rotation ($\beta_\mathrm{ini}\lesssim0.7\%$), $\beta_\mathrm{ini}$ can be approximated by a quadratic function of the initial angular speed $\Omega_\mathrm{ini}$ ($\beta_\mathrm{ini} \propto \Omega_\mathrm{ini}^2 [R_\mathrm{WD}^{(0)}]^3/M_\mathrm{WD}^{(0)}$), where $R_\mathrm{WD}^{(0)}$ and $M_\mathrm{WD}^{(0)}$ are the radius and mass of a non-rotating WD for each $M_\mathrm{DM}$ (a superscript ${(0)}$ stands for non-rotating case throughout this paper). The Keplerian velocity is decreased from $\sim10.9~\mathrm{rad/s}$ without DM admixture to $\sim5~\mathrm{rad/s}$ for $M_\mathrm{DM}=0.04~M_\odot$, due to the more extended radius. Limited by the Keplerian limit, the maximum $\beta_\mathrm{ini}$ is $\sim1.33\%$ for all our models, regardless of the amount of DM admixture. The initial WD mass is increased by at most $0.06~M_\odot$ in the near-Keplerian rotation case, while it is decreased when DM is admixed, by $0.2~M_\odot$ for $M_\mathrm{DM}=0.04~M_\odot$. The increment of $M_\mathrm{WD}$ due to rotation can be parametrized by 
\begin{equation} \label{eq:mwd_ini}
	M_\mathrm{WD} =  [1+(3.04\pm 0.02)\times\beta_\mathrm{ini}]M_{\mathrm{WD}}^{(0)},
\end{equation}
where $M_{\mathrm{WD}}^{(0)}$ is 1.44, 1.41, 1.37, 1.33, 1.28$~M_\odot$ for $M_\mathrm{DM}=0,~0.01,~0.02,~0.03,~\text{and}~0.04~M_\odot$, respectively.

\begin{deluxetable*}{ccccccc}[t!]
	\tablecaption{Global parameters of DM admixed rotating WDs. \label{tab:init_model_R5}}
	\tablecolumns{7}
	\tablenum{1}
	\tablewidth{0pt}
	\tablehead{
		\colhead{Model} &
		\colhead{$M_\mathrm{DM}~[M_\odot]$} &
		\colhead{$M_\mathrm{NM}~[M_\odot]$} & \colhead{$J~[10^{50}\mathrm{erg\cdot s}]$} & \colhead{$\beta_\mathrm{ini}~[\mathrm{\%}]$} &
		\colhead{$R_\mathrm{e}~[\mathrm{km}]$} &
		\colhead{$R_\mathrm{p}/R_\mathrm{e}$} 
	}
	\startdata
	R5-DM0 & 0    & 1.458 &  0.09   &    0.25   &  849  &  0.956 \\
	R5-DM1 & 0.01 & 1.416 &  0.11   &    0.33   &  934  &  0.942 \\
	R5-DM2 & 0.02 & 1.374 &  0.13   &    0.46   & 1043  &  0.928 \\
	R5-DM3 & 0.03 & 1.333 &  0.16   &    0.71   & 1243  &  0.896 \\
	R5-DM4 & 0.04 & 1.303 &  0.25   &    1.33   & 1809  &  0.726 \\
	\enddata
	\tablecomments{All models have a central baryonic matter density of $5\times10^{10}~\mathrm{g/cm}^3$ and initial angular velocity of $5~\mathrm{rad/s}$. $M_\mathrm{DM}$ and $M_\mathrm{NM}$ are the masses of dark matter and baryonic matter components; $J$ is the total angular momentum; $\beta_\mathrm{ini}$ is the initial ratio of rotational energy to gravitational energy; $R_\mathrm{e}$ and $R_\mathrm{p}$ are the equatorial and polar radii of the white dwarfs, respectively. \added{We extend this table to all our models in Tables~\ref{tab:all_models} and \ref{tab:beta_seris} of Appendix~\ref{app:extend}.}}
\end{deluxetable*}

\subsection{Hydrodynamics}
To follow the collapse of a white dwarf to PNS and the subsequent bounce and post-bounce dynamics, we solve the two-dimensional Euler equations of NM assuming axisymmetry \citep{2015MNRAS.454.1238L}:
\begin{eqnarray}
\nonumber \partial_t \rho + \nabla  \cdot (\rho \vec{v}) &=& 0,  \\ 
\partial_t (\rho\vec{v}) +  \nabla \cdot (\rho \vec{v}\vec{v}) + \nabla P &=& -\rho \nabla \Phi, \\ 
\nonumber \partial_t (\tau) + \nabla \cdot [(\tau+P)\vec{v}] &=& -\rho \vec{v}\cdot \nabla \Phi. 
\end{eqnarray}
Here $\tau=\rho(\epsilon+\frac{1}{2}v^2)$ is the total energy density, where $\epsilon$ is the specific internal energy and $\vec{v}$ is the fluid velocity. Our code utilizes a fifth-order shock capturing scheme Weighted-Essentially-Non-Oscillation  \cite[WENO;][]{1994JCoPh.115..200L} for the spatial discretization and 5-stage third-order Runge-Kutta scheme for time integration. As a first step, the DM admixture is assumed stationary and affects the dynamics of NM only through its gravity. The dynamics of DM accompanying AIC is an interesting problem and will be our future work.  We used a grid setup similar to that employed by the \texttt{FORNAX} code \citep{2016ApJ...831...81S}, which has a uniform resolution of $370~\mathrm{m}$ in the central $70~\mathrm{km}$ and becomes logarithmically increasing in the outer part ($\sim5~\mathrm{km}$ near the WD surface), and 45 angular grids are used in the quarter from the polar to equatorial planes. Simulations with finer resolutions in both radial and angular directions were performed and showed convergent results of the GW waveforms (Appendix~\ref{app:converge}). The hydrodynamic equations need to be closed with a gravity solver for $\Phi$ and EOS, which together with other microphysics inputs are described in the remaining part of this section. 

In our Newtonian hydrodynamics modeling, we mimic the relativistic gravity effects by modifying $\Phi$ and its derivative following the Case A formula presented in \cite{2006A&A...445..273M}, which has been shown to give reasonable agreements of the central density evolution and GW waveform compared to general relativistic (GR) simulations \citep[CFC+ approximation,][]{2002A&A...393..523D} for slowly rotating CCSNe. This implementation is widely used in many recent CCSN simulations \citep[e.g.~][]{2019MNRAS.482..351V}, even in the cases of black hole formation and failed supernovae \citep{2018ApJ...857...13P}. 

There are still large uncertainties in the nuclear matter EOS at high density for modeling CCSNe and NSs \citep{2017RvMP...89a5007O}. It has been explored extensively in \cite{2017PhRvD..95f3019R} for CCSN simulations, and we have tested the difference in results for AIC without DM admixture using 3 typical EOSs in Appendix \ref{app:EOS}. The EOS provided by \cite{1991NuPhA.535..331L} with compressibilty $K=220~\mathrm{MeV}$ (\texttt{LS220}) is selected for NM for our  discussion on the differences between AICs with different $M_\mathrm{DM}$. This choice is mainly because \texttt{LS220} has been widely used in many CCSN simulations and it agrees reasonably well with nuclear experiments and the measured masses and radii of NSs. Particularly, \texttt{LS220} was used in several recent studies of GW signals from CCSNe with multi-dimensional simulations \cite[e.g.][]{2013ApJ...779L..18C,2018ApJ...861...10M,2019MNRAS.486.2238A}. To use the finite temperature EOSs, we impose the same parameterized temperature profile to the initial models as \cite{2006ApJ...644.1063D}
\begin{equation}
T = T_{\rm c} [\rho/\rho_{\rm c,ini}]^{0.35},
\end{equation}
where the initial central temperature $T_{\rm c}$ is set to be 5~GK.

To trigger and follow the collapse of WDs, we use the parametrized electron capture scheme \citep{2005ApJ...633.1042L}, in which $Y_e$ is a function of NM density before the core bounce and neutrino pressure is included only in the trapping regime ($\rho>2\times10^{12}~\mathrm{g/cm}^3$). The $\rho-Y_e$ relation is obtained from the central trajectory of a spherically symmetric simulation of AIC with the open-source code \texttt{GR1D} \citep{2015ApJS..219...24O}, which has included a two-moment neutrino transport scheme and all the important emission and scattering reactions between neutrinos and NM.\footnote{We used Newtonian hydrodynamics with the effective GR potential in the \texttt{GR1D} simulation. The $\rho-Y_e$ relation is not affected by different treatments of the GR effect.} Some important \texttt{GR1D} results are presented in Appendix \ref{app:GR1D}. After the core bounce, no further deleptonization is included and $Y_e$ is simply advected. Since the post-bounce phase is evolved without neutrino transport, we present results mostly in the early post-bounce phase ($\lesssim15~\mathrm{ms}$ after bounce) during which neutrino leakage has a very small effect on the evolution \citep{2012PhRvD..86b4026O}. 

\subsection{Extraction of gravitational waves}
For our Newtonian hydrodynamical simulations, we utilize the standard quadrupole formula in the slow-motion and weak-field approximations to extract the GW strain $h_+$ from the simulations \citep{1991A&A...246..417M}
\begin{equation}
h_+ = \frac{3}{2} \frac{G}{Dc^4} \ddot{I}_\mathrm{zz} \sin^2\theta,
\end{equation}
where the source is placed at a distance of $D$ and orientation angle of $\theta$, and $\ddot{I}_\mathrm{zz}$ is the second time derivative of the mass-density quadrupole moment. While there are variants for performing the time derivative to minimize numerical noise \citep{1990ApJ...351..588F}, we found convergence among them by improving the accuracy of integration and optimizing the recording time-steps. 

\section{Results} \label{sec:results}
In this section, we present the major results from our hydrodynamical simulations of AIC of DM admixed rotating WDs. Section~\ref{subsec:hydro} describes how DM affects the collapse dynamics and properties of the inner core at the time of bounce ($t_\mathrm{b}$). Then we present the waveforms of the emitted GWs with various rotational speeds and $M_\mathrm{DM}$, as well as the detection prospect in Section~\ref{subsec:gw}. In Section~\ref{subsec:DM_signal}, we further analyze the GW signals and dig out the  imprinted information, especially on how to break the degeneracy between $M_\mathrm{DM}$ and the rotation rate.

\subsection{Dynamics} \label{subsec:hydro}
\begin{figure*}[t]
	\plottwo{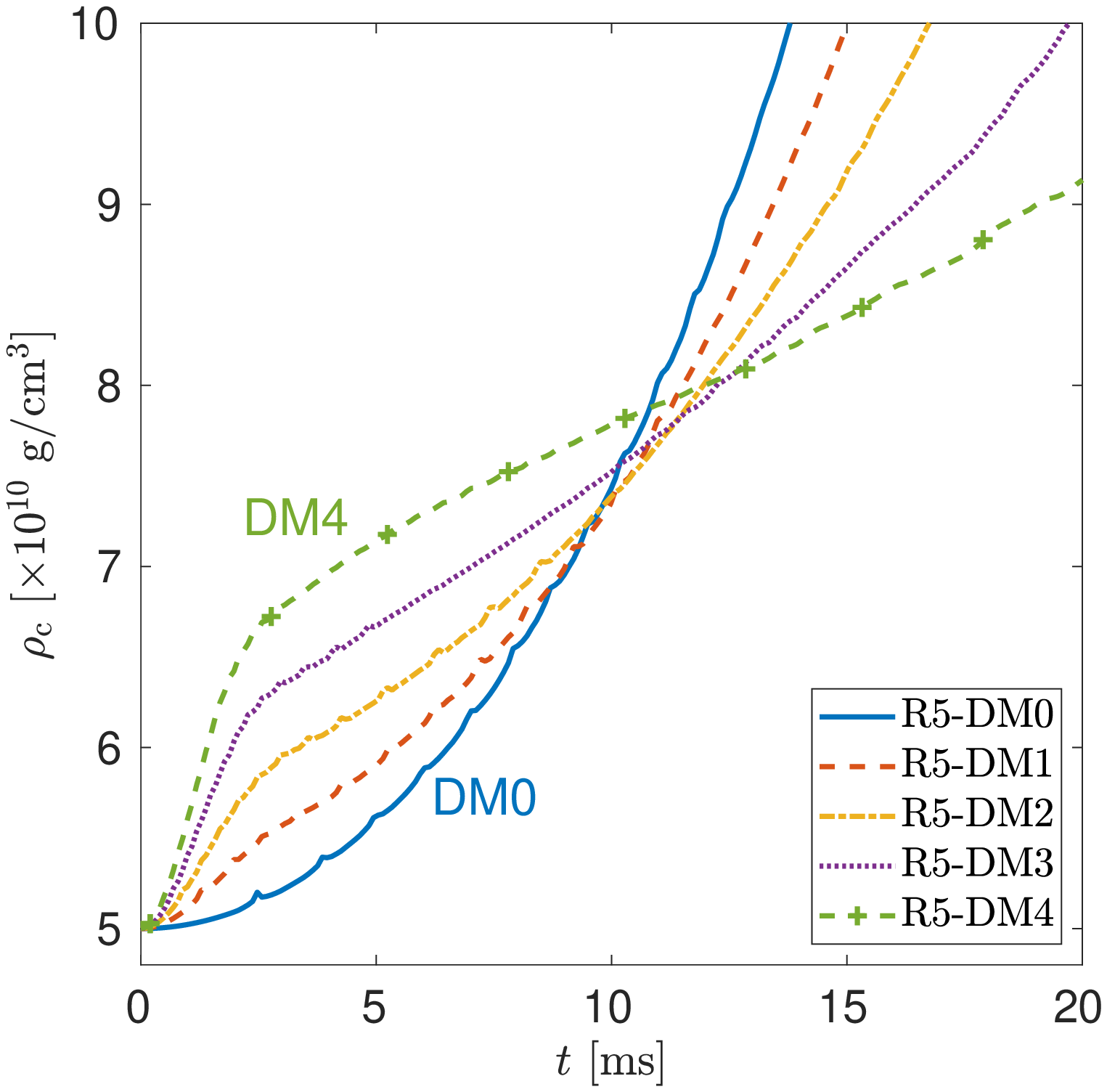}{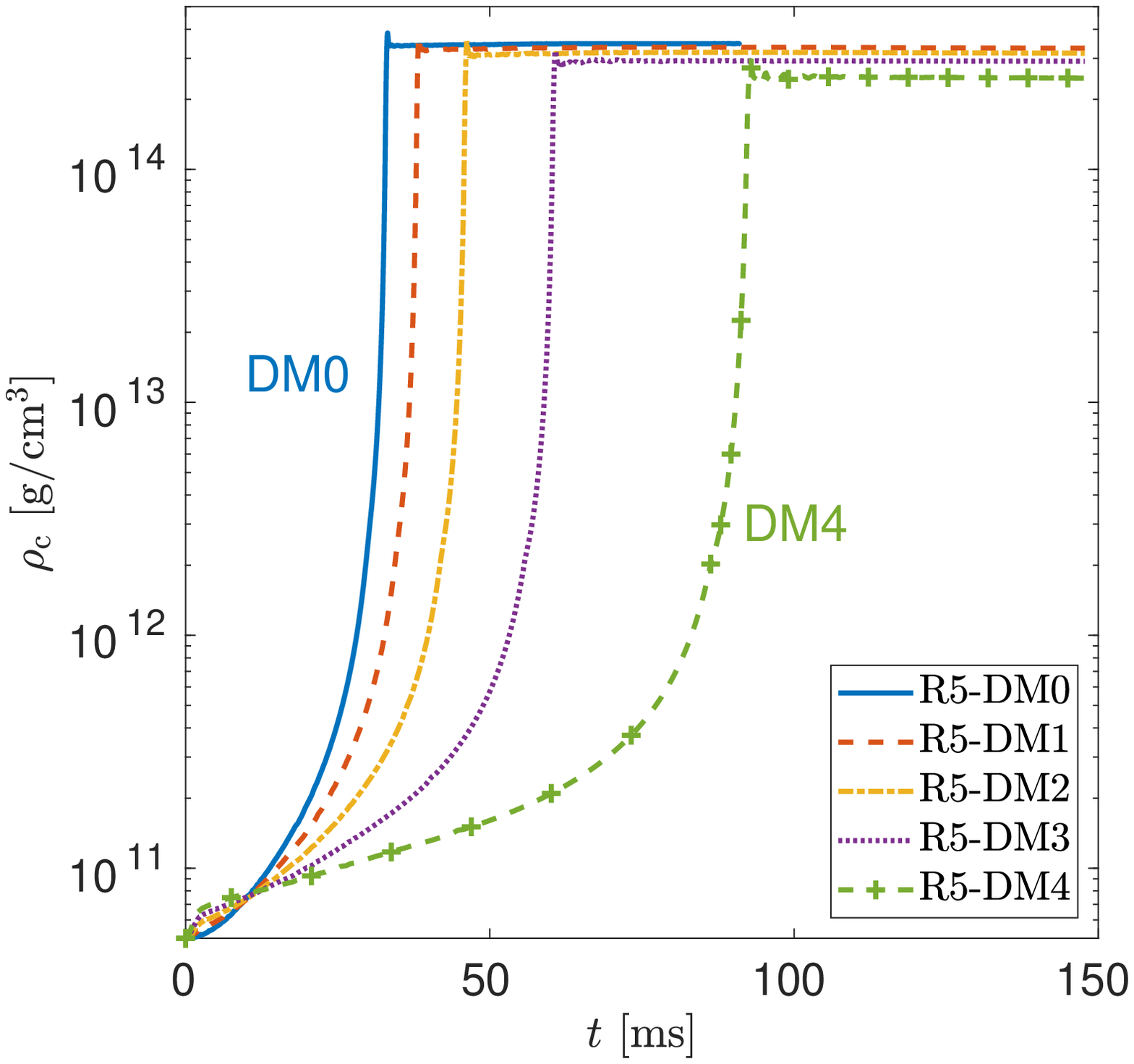}
	\caption{Central density evolution for the first $20~\mathrm{ms}$ (left) and entire time interval (right) of AIC simulations starting from models listed in Table \ref{tab:init_model_R5}. \label{fig:rhoc}}
\end{figure*}

AIC without DM admixture has been studied assuming spherical symmetry \citep{1999ApJ...516..892F} and axisymmetry \citep{2006ApJ...644.1063D, 2010PhRvD..81d4012A}, and we summarize the essential features of its dynamics here. A WD is supported by electron degeneracy pressure, and when it reaches the effective $M_\mathrm{Ch}$ this pressure support is reduced as electrons with high Fermi energy are captured by nuclei. The subsequent temporal evolution (Fig.~\ref{fig:rhoc}) can be divided into 3 phases similar to the canonical CCSNe \citep{2012ARNPS..62..407J}: infall, plunge and bounce, and ringdown. During the infall phase, $\rho_\mathrm{c}$ rises slowly and the WD is separated into two parts, a homologously collapsing inner core ($v_r\propto r$) and a supersonically collapsing outer core, which are in loose contact with each other. As $\rho_\mathrm{c}$ rises to above the nuclear saturation density ($\rho_\mathrm{sat}\simeq2.7\times10^{14}~\mathrm{g/cm}^3$), the inner core overshoots its equilibrium configuration and then sends out a hydrodynamic bounce shock, which turns into an accretion shock as its kinetic energy is lost due to disintegration of heavy nuclei and neutrino emissions \citep{2012ARNPS..62..407J}. Following \cite{2005ApJ...633.1042L}, the bounce time $t_\mathrm{b}$ is defined to be the instant when entropy per baryon at the edge of the inner core exceeds $3~k_\mathrm{B}$, which signifies the launch of the bounce shock.  

Fig.~\ref{fig:rhoc} shows the $\rho_\mathrm{c}$ evolution for the models listed in Table \ref{tab:init_model_R5}. In the early infall phase, as the central gravitational potential well is deeper for a larger $M_\mathrm{DM}$, models with more admixed DM show a faster contraction during the first $\sim5~\mathrm{ms}$ (left panel of Fig.~\ref{fig:rhoc}). But as DM admixture decreases $M_\mathrm{WD}$ and enlarges its size, the surge of $\rho_\mathrm{c}$ and bounce time $t_\mathrm{b}$ are delayed, and the central density at $t_\mathrm{b}$, $\rho_\mathrm{c,b}$, is decreased for a larger $M_\mathrm{DM}$ (right panel of Fig.~\ref{fig:rhoc}). These effects are qualitatively the same as found in non-rotating models \citep{2019LeungAIC}.

\begin{figure}[t]
	\vskip 1cm
	\plotone{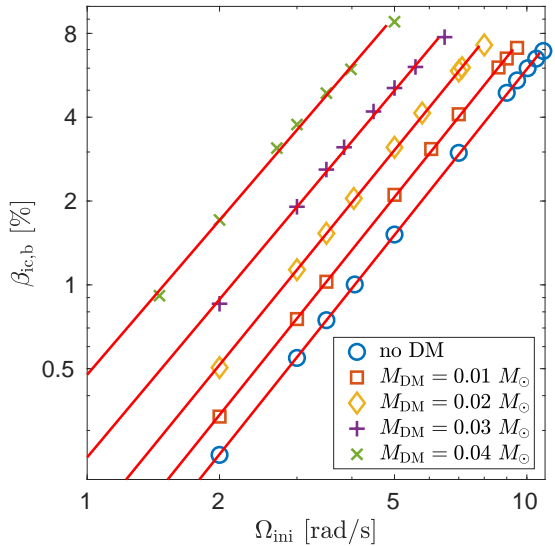}
	\caption{The $\beta$ parameter of the inner core at $t_\mathrm{b}$, $\beta_\mathrm{ic,b}$, as a function of $\Omega_\mathrm{ini}$ for different $M_\mathrm{DM}$.\label{fig:betab}}
\end{figure}

Faster rotation also delays the plunge and bounce phase, as well as decreases $\rho_\mathrm{c}$ through the centrifugal support. The $\beta$ parameter of the inner core at $t_\mathrm{b}$, $\beta_\mathrm{ic,b}$, is generally used as a measure for its rotation rate, and is found to strongly correlate with the emitted GW amplitude \citep{2008PhRvD..78f4056D,2010PhRvD..81d4012A}. In Fig.~\ref{fig:betab},  $\beta_\mathrm{ic,b}$ is plotted as a function of the initial angular velocity $\Omega_\mathrm{ini}$ for different $M_\mathrm{DM}$. Interestingly, $\beta_\mathrm{ic,b}$ shows a power-law relation to $\Omega_\mathrm{ini}$ ($\beta_\mathrm{ic,b}=C_\mathrm{b}\Omega_\mathrm{ini}^\alpha$) with an exponent $\alpha$ less than 2 that weakly depends on $M_\mathrm{DM}$ (the red lines in Fig.~\ref{fig:betab}) despite the highly nonlinear dynamical processes. The maximum value of $\beta_\mathrm{ic,b}$ is $\approx 9\%$ for $M_\mathrm{DM}=0.04~M_\odot$. This is below $\beta_\mathrm{dyn}\simeq0.25$ for developing the dynamical high-$\beta$ non-axisymmetric instability \citep{2007PhRvD..75d4023B}, while the elusive low-$\beta$ instability may still develop in full three-dimensional simulations \citep[see, e.g.][]{2007CoPhC.177..288C}. In this work, we only consider axisymmetric modeling.

\begin{figure*}[t!]
	\plottwo{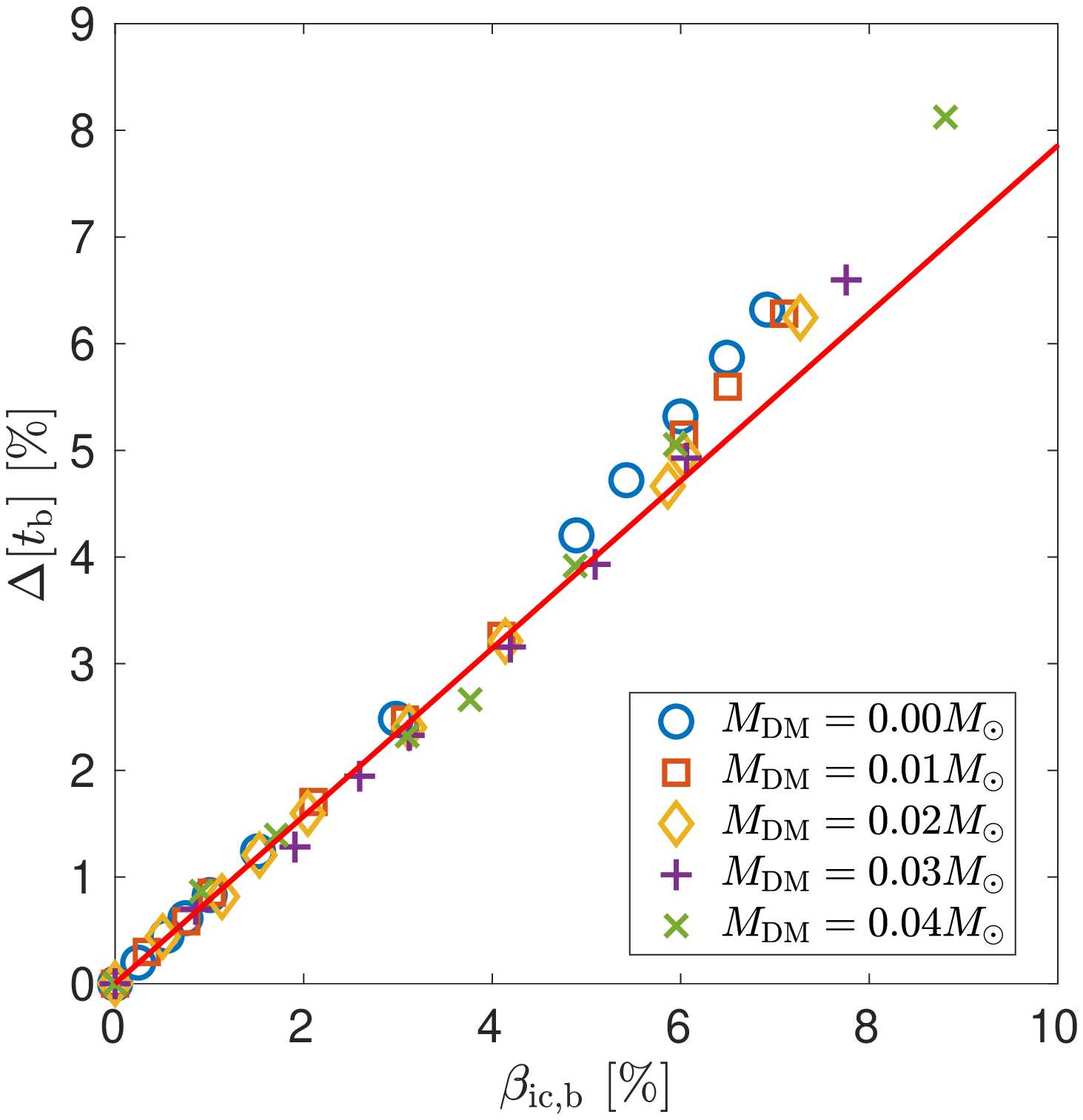}{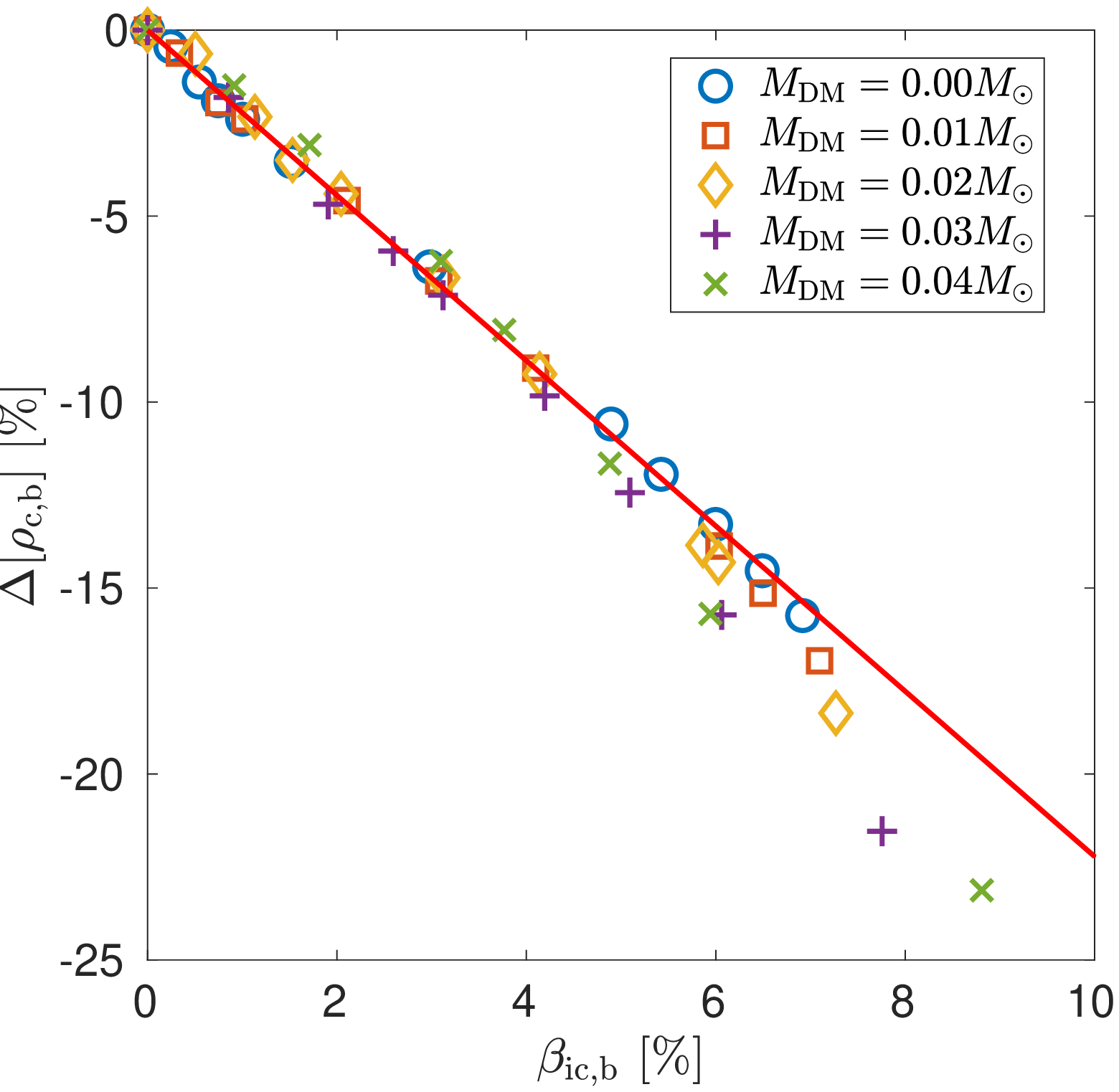}
	\caption{Relative differences of $t_\mathrm{b}$ (left) and $\rho_\mathrm{c,b}$  (right) between the rotating and non-rotating models with the same $M_\mathrm{DM}$ as a function of $\beta_\mathrm{ic,b}$. \label{fig:tb_rhoc}}
\end{figure*}

To disentangle the effects of DM admixture and rotation on the dynamics, we define the relative differences of a parameter $\mathcal{O}$, between the rotating and non-rotating models with the same $M_\mathrm{DM}$ as:
\begin{equation}
\Delta[\mathcal{O}] \equiv (\mathcal{O}-\mathcal{O}^{(0)}) / \mathcal{O}^{(0)}, \label{eq:relative_dif}
\end{equation} 
where $\mathcal{O}^{(0)}$ is for the non-rotating models. $\Delta[t_\mathrm{b}]$ and $\Delta[\rho_\mathrm{c,b}]$ are plotted as a function of $\beta_\mathrm{ic,b}$ in Fig.~\ref{fig:tb_rhoc}. It is clear that these relative differences are proportional to $\beta_\mathrm{ic,b}$ and have almost no explicit dependence on $M_\mathrm{DM}$, up to around $\beta_\mathrm{ic,b}\approx 6\%$ (red lines in both panels of Fig.~\ref{fig:tb_rhoc}). The offsets of $t_\mathrm{b}$ and $\rho_\mathrm{c,b}$ due to DM admixture can be calculated from the non-rotating models listed in Table~\ref{tab:NR}. Despite the decrease of $\rho_\mathrm{c,b}$ for a larger $M_\mathrm{DM}$ and $\beta_\mathrm{ic,b}$, all the models bounce at $\rho_\mathrm{c,b}>\rho_\mathrm{sat}$. This suggests that the uniform rotation of the progenitor WDs considered here is not too fast to result in a centrifugal bounce \citep{2008PhRvD..78f4056D}. 

\begin{deluxetable}{cccccc}[t!]
	\tablecaption{Parameters of the non-rotating models \label{tab:NR}}
	\tablecolumns{7}
	\tablenum{2}
	\tablewidth{0pt}
	\tablehead{
		\colhead{Model} &
		\colhead{$M_\mathrm{DM}$  } &
		\colhead{$t_\mathrm{b}^{(0)}$} &
		\colhead{$\rho_\mathrm{c,b}^{(0)}$} & 
		\colhead{$M_\mathrm{ic,b}^{(0)}$} &
		\colhead{$M_\mathrm{PNS}^{(0)}$} \\
		\colhead{ } &
		\colhead{$[M_\odot]$  } &
		\colhead{$[\mathrm{ms}]$} &
		\colhead{$[\mathrm{10^{14} g/cm^3}]$} & 
		\colhead{$[M_\odot]$} &
		\colhead{$[M_\odot]$}
	}
	\startdata
	R0-DM0 & 0    & 32.8 &  3.99   &    0.558  &  1.26 \\
	R0-DM1 & 0.01 & 37.7 &  3.90   &    0.543  &  1.21 \\
	R0-DM2 & 0.02 & 45.2 &  3.80   &    0.528  &  1.15 \\
	R0-DM3 & 0.03 & 58.6 &  3.69   &    0.510  &  1.07 \\
	R0-DM4 & 0.04 & 86.1 &  3.55   &    0.490  &  0.99 \\	\enddata
	\tablecomments{The superscript ${(0)}$ denotes non-rotating models. \added{We extend this table to all our models in Tables~\ref{tab:all_models} and \ref{tab:beta_seris} of Appendix~\ref{app:extend}. }}
\end{deluxetable}

\begin{figure}[t!]
	\plotone{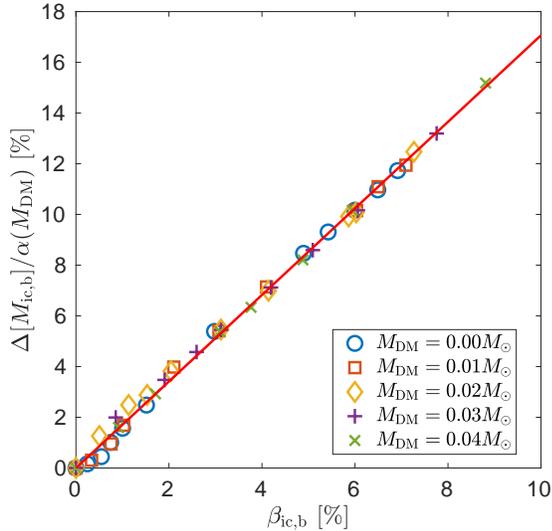}
	\caption{Same as Fig.~\ref{fig:tb_rhoc}, but for the relative differences of $M_\mathrm{ic,b}$ divided by $\alpha(M_\mathrm{DM})$.\label{fig:micb}}
\end{figure}

The mass of the inner core at the bounce, $M_\mathrm{ic,b}$, is another important parameter for AIC, which affects the strength of the bounce shock and also correlates with the GW amplitude. As listed in Table~\ref{tab:NR}, $M_\mathrm{ic,b}^{(0)}$ decreases with increasing $M_\mathrm{DM}$, in accordance with the smaller $M_\mathrm{WD}^{(0)}$ and $\rho_\mathrm{ic,b}^{(0)}$. $\Delta[M_\mathrm{ic,b}]$ increases linearly with $\beta_\mathrm{ic,b}$, and the increasing slope is smaller for a larger $M_\mathrm{DM}$. Fig.~\ref{fig:micb} shows the rescaled $\Delta[M_\mathrm{ic,b}]$ as a function of $\beta_\mathrm{ic,b}$, which demonstrates their linear correlation
\begin{equation}
\Delta[M_\mathrm{ic,b}] / \alpha(M_\mathrm{DM}) = (1.71\pm0.02) \times  \beta_\mathrm{ic,b}, \label{eq:mic_b}
\end{equation} 
where the denominator:
\begin{equation}
\alpha(M_\mathrm{DM})=3.24-[1-11.6(M_\mathrm{DM}/M_\odot)]^{-1},\label{eq:scaG_h2}
\end{equation} 
takes into account the $M_\mathrm{DM}$ dependence. This equation will be used in Section~\ref{subsec:DM_signal} for inferring $M_\mathrm{DM}$ from the GW amplitudes.

\subsection{Gravitational waves} \label{subsec:gw}
Rotating stellar collapses are expected to emit strong burst GWs and have been investigated for decades (see the review by \cite{2009CQGra..26f3001O}). Detailed hydrodynamical simulations of rotating CCSNe \citep{2008PhRvD..78f4056D} and AICs \citep{2010PhRvD..81d4012A} have shown that they emit GWs with generic waveforms. Here we study the dependence of the GW waveform on a new degree of freedom, $M_\mathrm{DM}$.

\begin{figure}[t!]
	\plotone{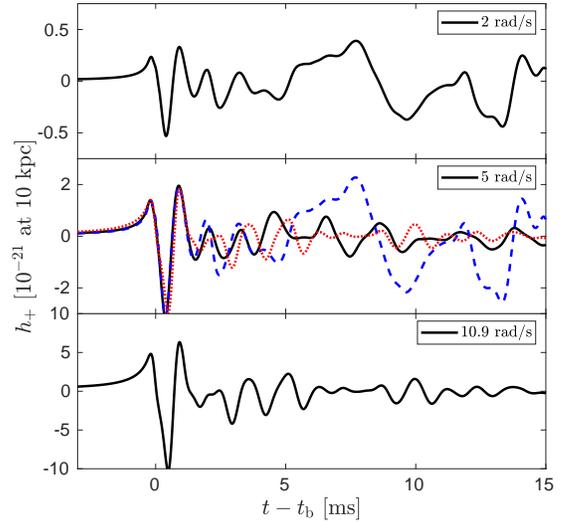} 
	\caption{GW waveform of AIC without DM admixture starting for the slowest~(upper panel), moderately~(middle panel), and fastest~(lower panel) rotating WDs. In the middle panel, the blue dashed~(red dotted) curve is the same as the upper~(lower) panel but multiplied with a constant factor, in order to illustrate the genericity of GW signals around $t_\mathrm{b}$. \label{fig:gwnodm}}
\end{figure}

Firstly, the GW waveforms of 3 normal rotating AIC models ($M_\mathrm{DM}=0$) are given in Fig.~\ref{fig:gwnodm}. They represent our slowest, moderately, and fastest rotating WD models. The waveform type is different from that in \cite{2010PhRvD..81d4012A} according to the classification in \cite{2009CQGra..26f3001O}, and this will be discussed in Section~ \ref{subsec:dis_ecap}. Similar to the CCSNe \citep{2008PhRvD..78f4056D}, the waveforms from AIC models in this study are Type I (pronounced spikes around $t_\mathrm{b}$ associated with core bounce induced by stiffening of the nuclear EOS, followed by ``ring-down" oscillations) and can be divided into two sub-groups: the slow rotating models (upper panel in Fig.~\ref{fig:gwnodm}) show significant contributions from prompt convection after $\sim5~\mathrm{ms}$ post-bounce as long period oscillations, while fast rotating models (middle and lower panels in Fig.~\ref{fig:gwnodm}) show dominantly post-bounce ring-down signals only. None of our models displays centrifugal bounce since the maximum $\beta_\mathrm{ic,b}$ is quite small ($<9\%$), and $\rho_\mathrm{c,b}$ is always above $\rho_\mathrm{sat}$ for the initial WDs with uniform rotation. In the middle panel of Fig.~\ref{fig:gwnodm}, the dashed and dotted curves are those in the upper and lower panels but multiplied by a constant factor. The excellent match of the spikes around $t_\mathrm{b}$ shows the genericity of the waveforms, and this feature will be further analyzed in Section~\ref{subsec:DM_signal}.

\begin{figure}[t!]
	\plotone{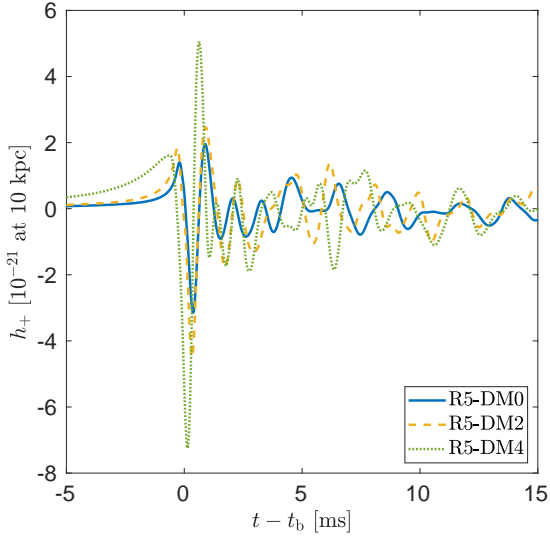}
	\caption{GW waveforms of AICs starting from DM admixed rotating WDs, with the same initial angular velocity $\Omega_\mathrm{ini}=5~\mathrm{rad/s}$ but different $M_\mathrm{DM}$. \label{fig:gwr4dm}}
\end{figure}

We then show the waveforms from models with the same $\Omega_\mathrm{ini}$ of $5~\mathrm{rad/s}$ but different $M_\mathrm{DM}$ in Fig.~\ref{fig:gwr4dm}. With DM admixture, the waveform shape still belongs to the Type I category. For the same $\Omega_\mathrm{ini}$ and $\rho_\mathrm{c}$, a larger $M_\mathrm{DM}$ leads to a larger $\beta_\mathrm{ini}$ (Fig.~\ref{fig:init}) and $\beta_\mathrm{ic,b}$ (Fig.~\ref{fig:betab}). This results in some enhancement of the GW emission, especially amplitudes of the three spikes around the time of bounce. 

\begin{figure}[t!]
	\plotone{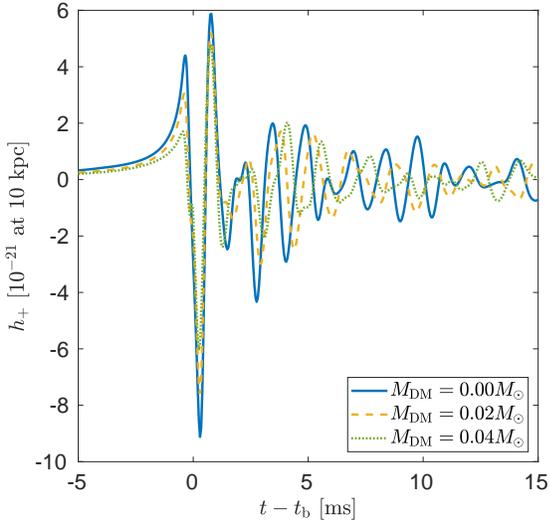}
	\caption{Same as Fig.~\ref{fig:gwr4dm}, but for models with different $M_\mathrm{DM}$ but reaching the same $\beta_\mathrm{ic,b}\sim6\%$.  \label{fig:gwbetadm}}
\end{figure}

Results in \cite{2008PhRvD..78f4056D} and \cite{2010PhRvD..81d4012A} suggest that the GW amplitude has a strong correlation with $\beta_\mathrm{ic,b}$, and so $\beta_\mathrm{ic,b}$ can in principle be inferred from GW observations. In Fig.~\ref{fig:gwbetadm} we pick models with the same $\beta_\mathrm{ic,b}$ ($\simeq6~\%$) but different $M_\mathrm{DM}$. For this set of models, the GW amplitude around the bounce decreases significantly as $M_\mathrm{DM}$ increases. For example, the peak before the bounce decreases by 2.5 times for $M_\mathrm{DM}$ from 0 to $0.04~M_\odot$. This is related to the decrement of $\rho_\mathrm{c}$ (Fig.~\ref{fig:tb_rhoc}) and $M_\mathrm{ic}$ (Fig.~\ref{fig:micb}), thus less compact core, for a larger $M_\mathrm{DM}$. We will analyze these changes quantitatively in Section~\ref{subsec:DM_signal} to disentangle the effects of DM admixture and rotation rate.

\begin{figure}[t!]
	\plotone{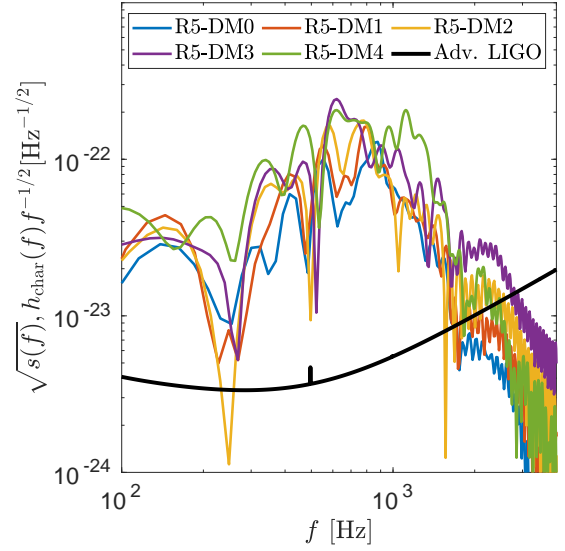}
	\caption{Characteristic strain spectra for selected models listed in Table \ref{tab:init_model_R5} compared with the noise spectrum of Adv. LIGO. $\sqrt{s(f)}$ is the one-sided noise amplitude spectral density of a GW detector \citep[black solid: LIGO noise curve, in file aLIGODesign.txt from][]{LIGO_v5}.\label{fig:hchar}}
\end{figure}

For the detection prospect, the characteristic amplitude $h_\mathrm{char}$ of GWs from DM admixed AICs is compared to the noise spectra of LIGO. Following \cite{2009ApJ...707.1173M}, the dimensionless characteristic amplitude can be calculated by
\begin{equation}
h_\mathrm{char} = \frac{1}{D} \sqrt{\frac{2G}{\pi^2 c^3} \frac{\mathrm{d}E_\mathrm{GW}}{\mathrm{d}f}},
\end{equation}
where $\frac{\mathrm{d}E_\mathrm{GW}}{\mathrm{d}f}$ is the GW spectral energy density,
\begin{equation}
\frac{\mathrm{d}E_\mathrm{GW}}{\mathrm{d}f} = \frac{4}{15}\frac{c^3}{G}(2\pi f)^2 D^2|\tilde{h}_+|^2,
\end{equation}
and $\tilde{h}_+$ is the Fourier transform of GW strain $h_+$
\begin{equation}
\tilde{h}_+(f) = \int_{-\infty}^{\infty} h_+(t) e^{-2\pi ift}\mathrm{d}t. \label{eq:fourier}
\end{equation}
The Fourier transform includes signals only until $10~\mathrm{ms}$ post-bounce to avoid possible contribution from prompt convection, and the $h_\mathrm{char}$ spectra of selected AIC models with different $M_\mathrm{DM}$ listed in Table \ref{tab:init_model_R5} are plotted in Fig.~\ref{fig:hchar}, assuming that the AIC events are happening at $10~\mathrm{kpc}$ (within the Milky Way) from the detectors. The GWs have a broad frequency contribution from $\sim200~\mathrm{Hz}$ to $\sim1500~\mathrm{Hz}$ and several peaks in-between, which lie in the most sensitive detection band of LIGO. However, from binary population synthesis calculations, the Galactic AIC rate is expected to be $10^{-4}-10^{-3}~\mathrm{yr^{-1}}$ summing over all possible progenitor scenarios \citep{2018MNRAS.481..439W,2019MNRAS.484..698R}, which disfavors the detection of an AIC event. We estimated that with the proposed sensitivity of the Einstein Telescope \citep{2011CQGra..28i4013H}, the detection distance can be increased to $\sim1~\mathrm{Mpc}$, which will boost the detection possibility significantly.

\subsection{DM imprints in GW} \label{subsec:DM_signal}
The GW waveforms shown in Fig.~\ref{fig:gwnodm},~\ref{fig:gwr4dm} and \ref{fig:gwbetadm} have three dominant spikes around $t_\mathrm{b}$. We denote the amplitudes of these spikes as $h_1$ (positive, before $t_\mathrm{b}$), $h_2$ (negative, after $t_\mathrm{b}$) and $h_3$ (positive, after $t_\mathrm{b}$). Previous studies \citep{2017PhRvD..95f3019R} found that for CCSNe, $h_1$ and $\Delta h = h_1-h_2$ increase monotonically with increasing $\beta_\mathrm{ic,b}$ for $\beta_\mathrm{ic,b}\le7\%$, and this correlation has relatively weak dependence on the EOS and differential rotation. For our DM admixed models, the amplitudes also depend on $M_\mathrm{DM}$ and so there is a degeneracy between the two parameters, $M_\mathrm{DM}$ and $\beta_\mathrm{ic,b}$. 

\begin{figure*}[ht!]
	\plottwo{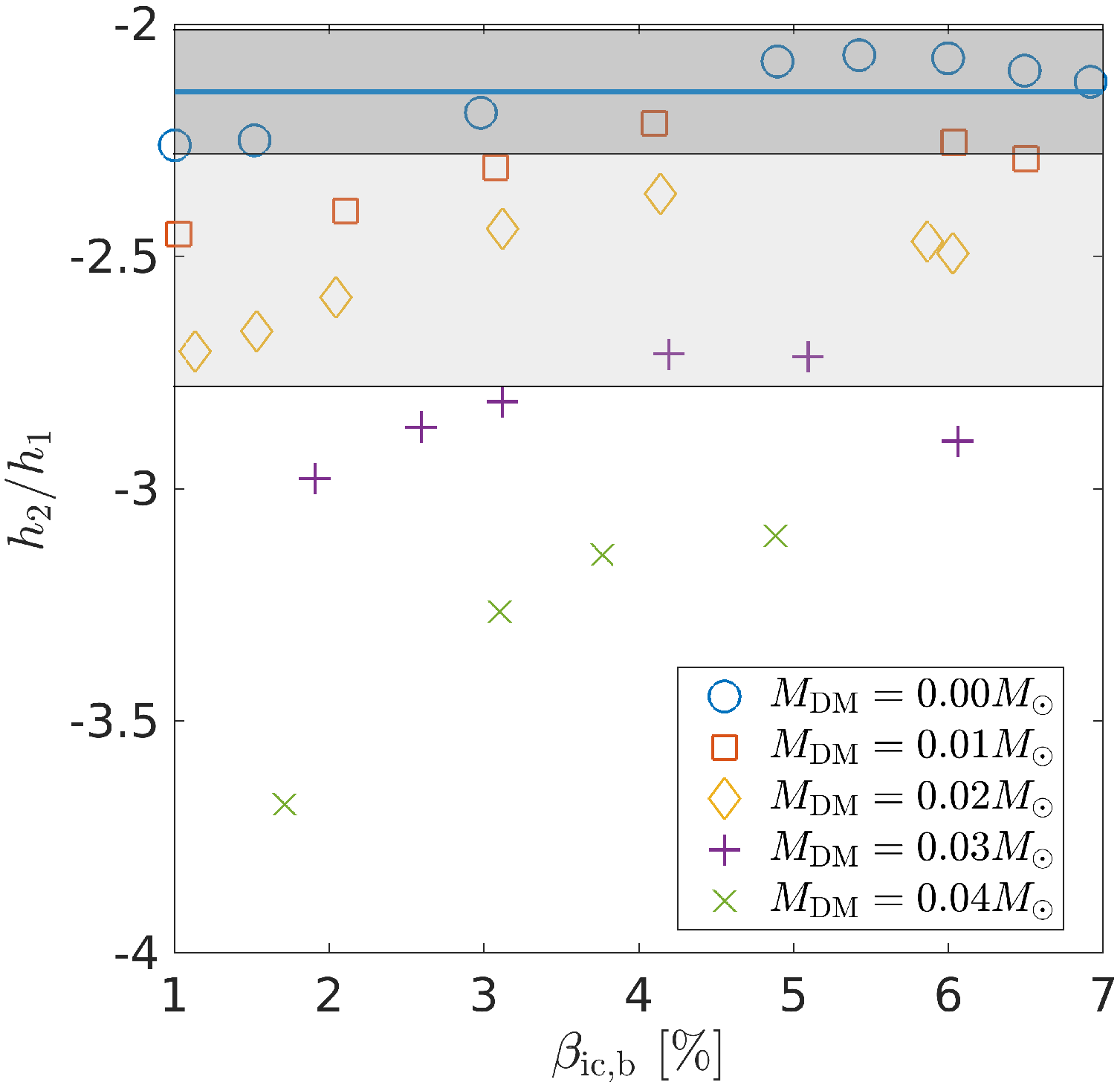}{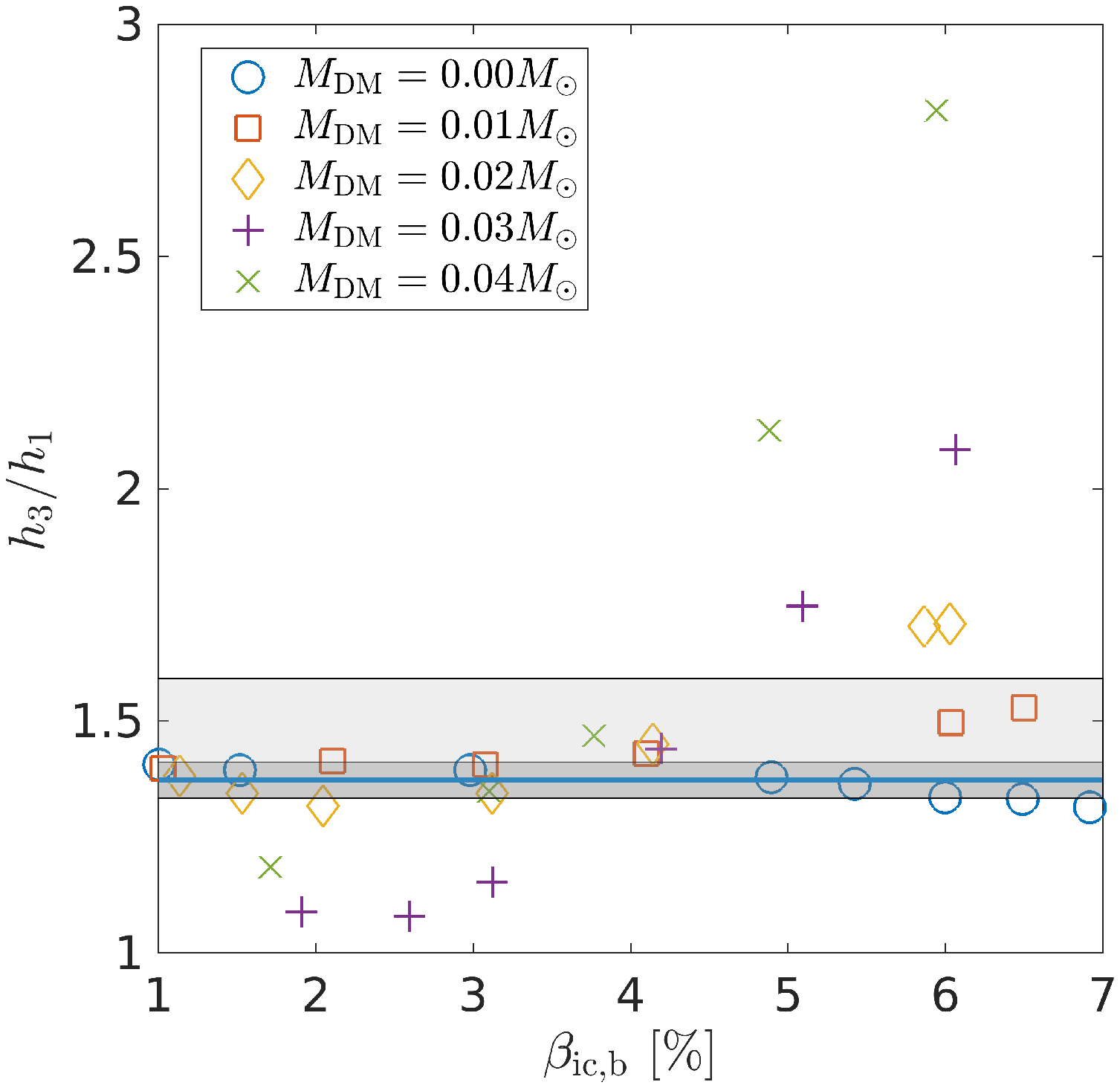}
	\caption{Ratios between the amplitudes of GW spikes ($h_{1,2,3}$ defined in the text) around $t_b$ as a function of $\beta_\mathrm{ic,b}$ for different $M_\mathrm{DM}$. In both panels, the thick shaded region represents the variation due to different $\beta_\mathrm{ic,b}$ and the light shaded region due to different EOSs. \label{fig:hratio}}
\end{figure*}

We plot the ratios of these peak amplitudes, i.e. $h_2/h_1$ and \replaced{$h_3/h_2$}{$h_3/h_1$}, as a function of $\beta_\mathrm{ic,b}$ for different $M_\mathrm{DM}$ in Fig.~\ref{fig:hratio}. Without DM admixture, the two ratios have relatively small variations for different $\beta_\mathrm{ic,b}$, with $h_2/h_1=-2.14\pm0.14$ and $h_3/h_1=1.37\pm0.04$. DM admixture breaks down this invariance. The absolute values and variation of $h_2/h_1$ for different $\beta_\mathrm{ic,b}$ are generally larger for a larger $M_\mathrm{DM}$. For $h_3/h_1$, the absolute value is smaller at $\beta_\mathrm{ic,b}\lesssim3\%$ and larger at $\beta_\mathrm{ic,b}\gtrsim4\%$ for a larger $M_\mathrm{DM}$. For a fixed $M_\mathrm{DM}$, $h_3/h_1$ is larger for faster rotation. Therefore, deviation of $h_2/h_1$ and $h_3/h_1$ from those of DM-absent models in a GW observation can indicate the presence of DM admixture. The relatively small spread of the ratios is also true for the CCSNe GW catalog provided by \cite{2017PhRvD..95f3019R} though the mean values of $h_2/h_1$ depend on the specific EOS and are between -2 and -3. The light shaded regions in Fig.~\ref{fig:hratio} represent the uncertainties introduced by different EOSs simulated in Appendix~\ref{app:EOS}. If the DM core has a mass $M_\mathrm{DM}\geq 0.03~M_\odot$, its existence can be inferred from the GW signals, despite our ignorance of the EOS. The presence of DM admixture with $M_\mathrm{DM}\leq 0.02~M_\odot$ can be inferred only if the EOS is better constrained.

\begin{figure*}[ht!]
	\plottwo{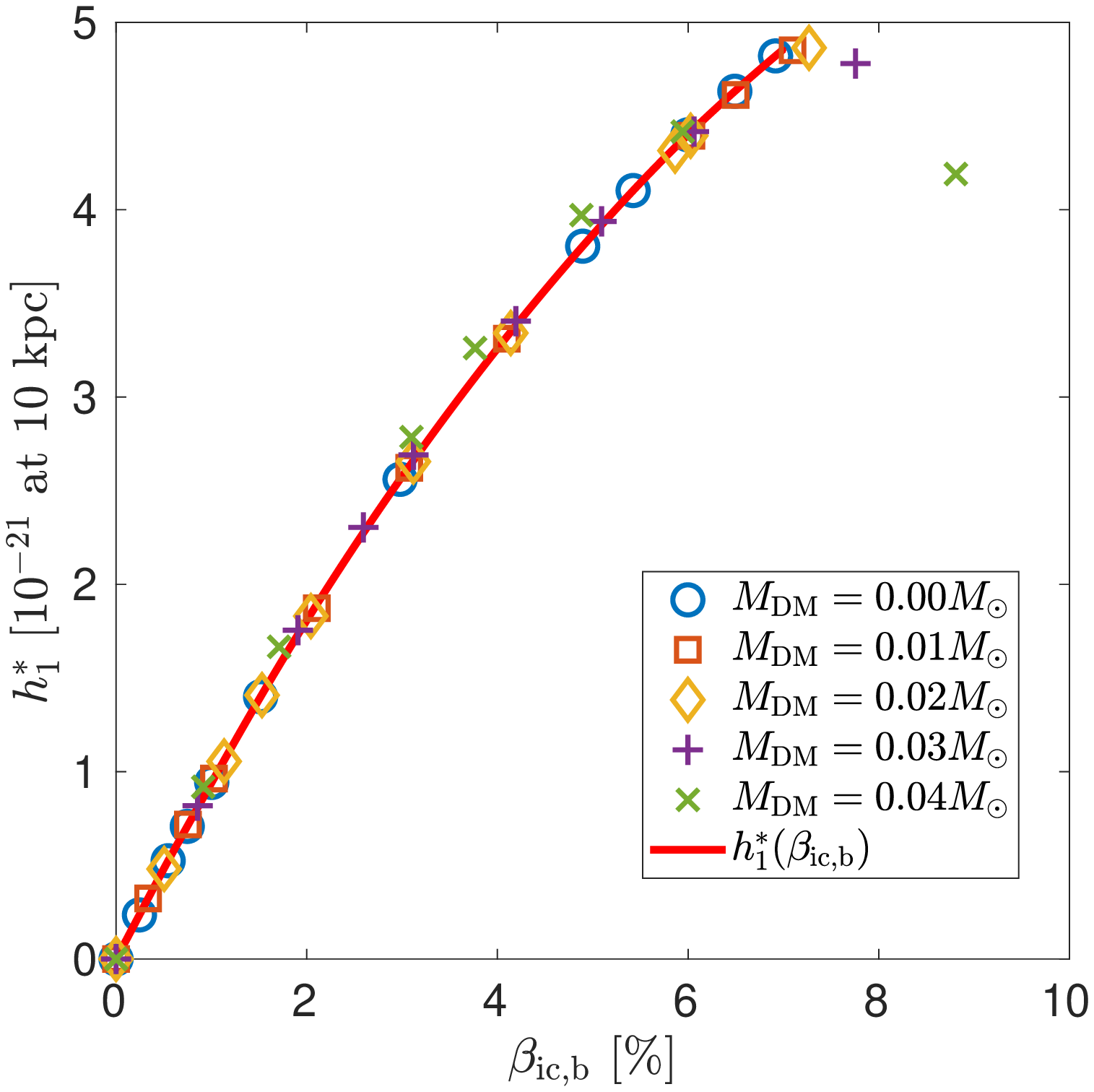}{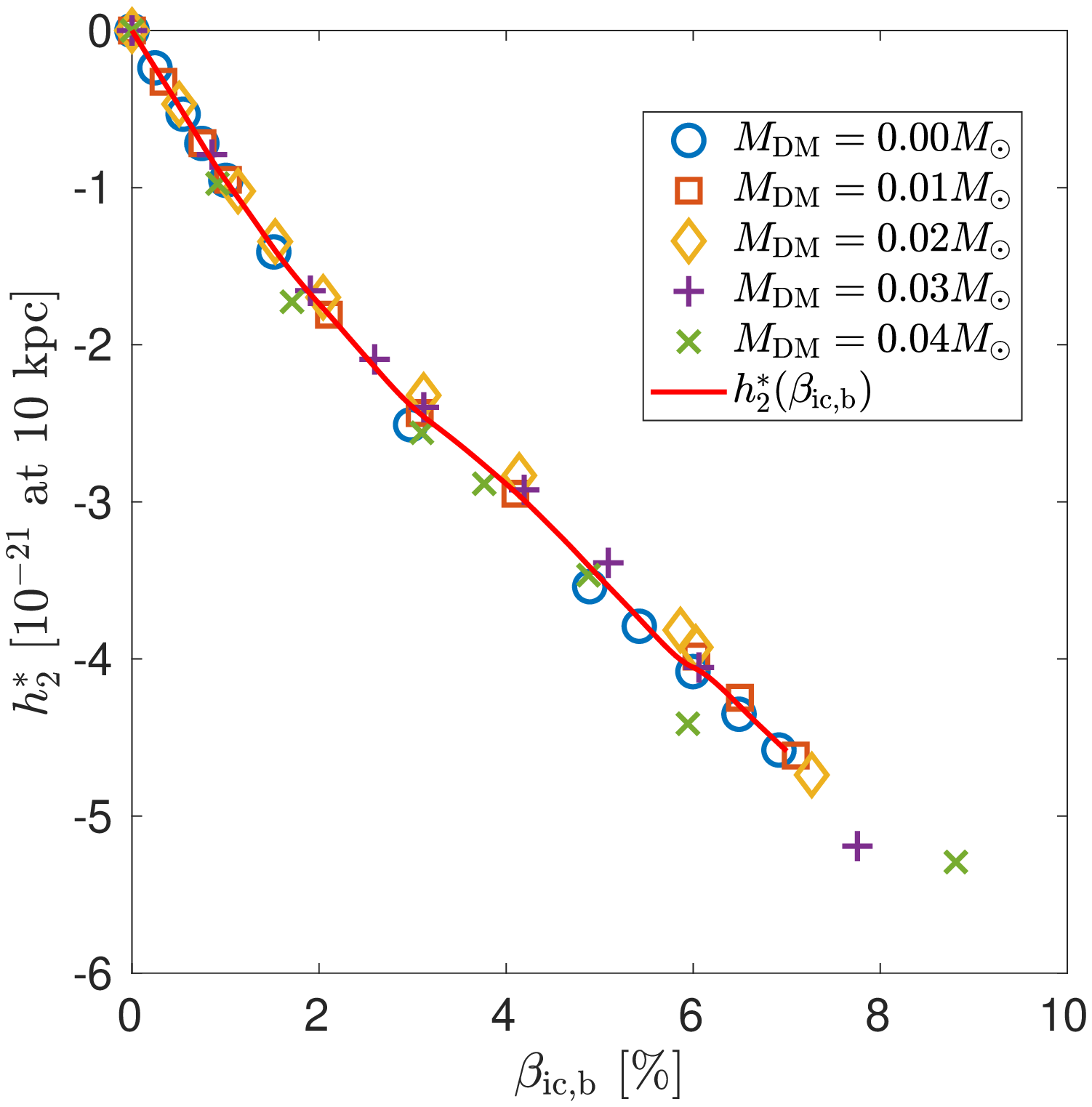}
	\caption{Same as Fig.~\ref{fig:hratio}, but for the rescaled values of $h_{1,2}$. The two red curves are interpolated from all the points at $\beta_\mathrm{ic,b}\le7\%$ with the cubic spline method. The rescaling of $h_{1,2}$ are given by Eqs.~\ref{eq:scaG_h1} and \ref{eq:h2_star}, respectively .\label{fig:hscaling}}
\end{figure*}

\added{If a GW event of AIC is detected by a GW detector such as LIGO, can we infer the DM admixing? The mismatch ($\mathcal{M}$) between two GW waveforms for a given GW detector is \citep{2011CQGra..28s5015R,2017PhRvD..95f3019R}
\begin{equation}
\mathcal{M}=1-\max_{t_{\rm A}}\bigg[ \frac{\langle h_a,h_b \rangle}{\sqrt{\langle h_a,h_a \rangle \langle h_b,h_b \rangle}}\bigg], \label{eq:mismatch}
\end{equation}
where the second term is the match between the two waveforms $h_a$ and $h_b$, maximized over the difference between their arrival times $t_{\rm A}$. The inner product ${\langle h_a,h_b \rangle}$ is calculated by
\begin{equation}
<h_a,h_b> = \int_{0}^{\infty} \frac{4\tilde{h}_a^*\tilde{h}_b}{s} \mathrm{d}f,
\end{equation}
where the Fourier spectrum $\tilde{h}$ is defined in Eq.~\ref{eq:fourier} and $s$ is the noise spectrum of the LIGO detector (the black line in Figure~\ref{fig:hchar}). Here the integration limit is from 100 Hz to 2000 Hz, which is the dominant frequency range for the GW waveforms in this work.

\begin{table}[t!]
	\centering
	\caption{\added{Mismatches ($\mathcal{M}$) between the GW waveforms with the same $\beta_\mathrm{ic,b}$, but without and with DM admixture. Both $\mathcal{M}$ and $\beta_\mathrm{ic,b}$ are in the unit of $\%$.} \label{tab:mismatch}}
	\begin{tabular}{|c|c|c|c|c|c|c|}
		\hline
		& \multicolumn{2}{c|}{$\beta_\mathrm{ic,b}\approx1\%$} & \multicolumn{2}{c|}{$\beta_\mathrm{ic,b}\approx3\%$} & \multicolumn{2}{c|}{$\beta_\mathrm{ic,b}\approx6\%$} \\ \hline
		$M_{\rm DM}~[M_\odot]$ & $\beta_\mathrm{ic,b}$ & $\mathcal{M}$ & $\beta_\mathrm{ic,b}$ & $\mathcal{M}$ & $\beta_\mathrm{ic,b}$ & $\mathcal{M}$ \\ \hline
		0      & 1.00  & 0   & 2.98 & 0  & 6.00 & 0 \\
		0.01 &1.03   & 8   & 3.08 & 7   & 6.03 & 5 \\
		0.02 & 1.13  & 14 & 3.12  & 20 & 6.03 &  19\\
		0.03 & 0.86 & 39   & 3.12  & 31 & 6.06 & 34\\
		0.04 & 0.91 & 35 & 3.10  & 39 & 5.94 &  47\\
		\hline
	\end{tabular}
\end{table}

In Table~\ref{tab:mismatch} we list the mismatches between the GW waveforms with the same $\beta_\mathrm{ic,b}$ ($\approx1\%,~3\%,~6\%$), but without and with DM admixture. Our results show that if the LIGO detector can distinguish two waveforms with $\mathcal{M}\gtrsim10\%$, then the DM admixture with $M_{\rm DM}\gtrsim 0.02~M_\odot$ can be inferred.
}

To retrieve the two parameters $M_\mathrm{DM}$ and $\beta_\mathrm{ic,b}$ from a GW observation, we further analyze the dependence of $h_{1,2}$ on $M_\mathrm{DM}$ and $\beta_\mathrm{ic,b}$ in Fig.~\ref{fig:hscaling}. The dependence of $h_1$ on $M_\mathrm{DM}$ can be removed by rescaling it:
\begin{equation} \label{eq:scaG_h1}
h_1^* \equiv h_1 / [1-15.36(M_\mathrm{DM}/M_\odot)].
\end{equation}
For $\beta_\mathrm{ic,b}\lesssim7\%$, $h_1^*$ follows a unified and monotonic increasing relation with $\beta_\mathrm{ic,b}$. The analysis of $h_2$ is more subtle. We found that $h_2$ increases linearly with $\Delta[M_\mathrm{ic,b}]$ (see Section~\ref{subsec:hydro} and Eq.~\ref{eq:relative_dif} for the definition), which when rescaled by $\alpha(M_\mathrm{DM})$ (Eq.~\ref{eq:scaG_h2}) is proportional to $\beta_\mathrm{ic,b}$ (Fig.~\ref{fig:micb}). Therefore,
\begin{equation}
h_2^* \equiv \frac{h_2}{3.24-[1-11.6(M_\mathrm{DM}/M_\odot)]^{-1}}  \label{eq:h2_star}
\end{equation}
also follows a unified and monotonic increasing relation with $\beta_\mathrm{ic,b}$. 

Using the universal relations, $h_1^*({\beta_\mathrm{ic,b}})$ and $h_2^*({\beta_\mathrm{ic,b}})$, in principle, $M_\mathrm{DM}$ and $\beta_\mathrm{ic,b}$ can be retrieved from accurate measurement of $h_{1,2}$ in a GW observation. In addition, the whole waveform including the post-bounce ring-down oscillations can confirm this determination.
One caution is that the microphysics inputs, such as $Y_e=Y_e(\rho)$ profile and EOS, may affect the exact functional forms of $h_{1,2}^*$. Fig.~\ref{fig:gw_eos} shows that $h_1$ hardly changes for different EOSs while $h_2$ could change by $\sim30\%$. \cite{2017PhRvD..95f3019R} showed that $\Delta h=h_1-h_2$ varies by $\sim30\%$ when the electron capture rate is scaled by 0.1 and 10. So a firm retrieval of $M_\mathrm{DM}$ awaits for better constrained microphysics inputs.

\section{Discussion} \label{sec:discuss}
\subsection{Pre-bounce electron capture} \label{subsec:dis_ecap}
\cite{2010PhRvD..81d4012A} has studied rotating AICs without DM admixture with general relativistic simulations using the CoCoNuT code \citep{2002A&A...393..523D}. They used the electron parametrization profile, $Y_e=Y_e(\rho)$, from AIC simulations with Multi-Group Flux-Limited Diffusion approximation for neutrino transport \citep{2006ApJ...644.1063D}, which resulted in a very low central $Y_e$ at the time of bounce ($\sim 0.19$ compared to $\sim 0.27$ in simulations of CCSNe with more accurate neutrino transport schemes). The small central $Y_e$ leads to a small mass of the homologous collapsing inner core ($M_\mathrm{ic,b}\simeq0.26~M_\odot$ for $\beta_\mathrm{ic,b}\le 10\%$) and a subdominant negative spike after bounce in GW emission, belonging to the Type III signal \citep{2009CQGra..26f3001O}. In our study, the parametrization profile  obtained from \texttt{GR1D} simulations (Appendix \ref{app:GR1D}) is closer to those of standard CCSNe. The presented GW waveforms in Figs.~\ref{fig:gwnodm},~\ref{fig:gwr4dm} and \ref{fig:gwbetadm} are generically Type I. 

\begin{figure}
	\plotone{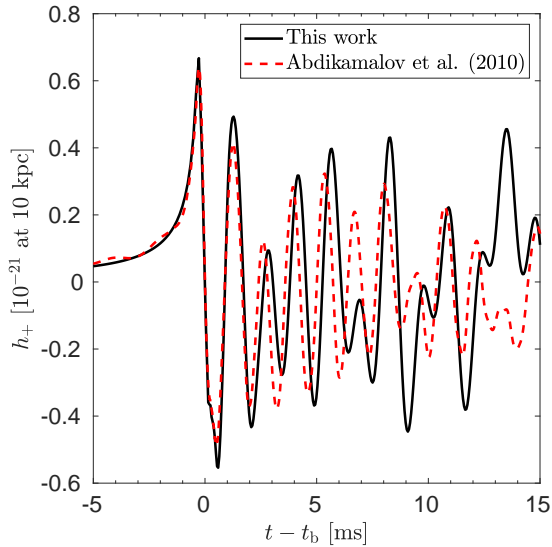}
	\caption{GW waveform (black solid line) from an AIC simulation with $\Omega_\mathrm{ini}=5~\mathrm{rad/s}$ and the same $Y_e=Y_e(\rho)$ profile as that used in \cite{2010PhRvD..81d4012A} compared with that from their GW catalog (red dashed line).\label{fig:gw_compare}}. 
\end{figure}

To check whether the different results obtained by us and \cite{2010PhRvD..81d4012A} are due to the microphysics employed or GR effect, we performed an AIC simulation with $\Omega_\mathrm{ini}=5~\mathrm{rad/s}$ and the same $Y_e=Y_e(\rho)$ profile as that used in \cite{2010PhRvD..81d4012A}, and the GW waveform is shown in Fig.~\ref{fig:gw_compare} compared to their result. The three spikes around $t_\mathrm{b}$ match quite well for the two waveforms, with $\sim10\%$ difference in peak amplitudes. Our results agree with the study by \cite{2019arXiv190109055P} who found a nearly identical bounce signal between CCSNe simulations with \texttt{CoCoNuT} and the Case A effective GR potential. The ring-down oscillations have different periods, which may be due to a discrepancy in the equilibrated PNS structure and/or grid resolution. As the numerical calculations of electron capture in the collapse phase is still uncertain \citep{2019ApJS..240...38N}, it would be interesting to study how this would generally affect our results. We hope to return to this issue in the future. Since our focus is the effects of DM admixture, the conclusions drawn from the bounce GW signals are expected not to be altered qualitatively.

\subsection{PNS mass} \label{subsec:mpns}
Apart from the signal around $t_b$, the post-bounce convective motion inside and above the newborn PNS also emits significant GWs and is under intensive investigation recently in the context of CCSNe \citep[e.g.~][]{2019ApJ...876L...9R,2019arXiv190109055P,2019arXiv190210048T}. An important emission mechanism is the g-mode oscillation at the surface of the PNS. \cite{2013ApJ...766...43M} found that the peak frequency of this g-mode GW is proportional to the PNS mass ($M_\mathrm{PNS}$). We use $M_\mathrm{PNS}$ at $50~\mathrm{ms}$ post-bounce to investigate the dependence of this signal on model parameters $M_\mathrm{DM}$ and $\beta_\mathrm{ic,b}$, where the PNS is defined as the core with $\rho\ge10^{11}~\mathrm{g/cm^3}$. Table~\ref{tab:NR} lists $M_\mathrm{PNS}^{(0)}$ for the non-rotating models and it decreases from $1.26~M_\odot$ without admixed DM to $0.99~M_\odot$ for $M_\mathrm{DM}=0.04~M_\odot$. For a larger $M_\mathrm{DM}$, the g-mode frequency is decreased by approximately $\sim20\%$ for $M_\mathrm{DM}=0.04~M_\odot$, compared to that without DM admixture. 

Faster rotation makes the PNS less compact and generally lighter through the centrifugal support. $\Delta[M_\mathrm{PNS}]$ (defined in Eq.~\ref{eq:relative_dif}) are shown in Fig.~\ref{fig:dmpns} as a function of $\beta_\mathrm{ic,b}$.  $M_\mathrm{PNS}$ decreases linearly with increasing $\beta_\mathrm{ic,b}$ and the decrement is $\sim10\%$ at $\beta_\mathrm{ic,b}\simeq8\%$. DM admixture makes $\Delta[M_\mathrm{PNS}]$ larger for the same $\beta_\mathrm{ic,b}$, but this effect is not monotonic for increasing $M_\mathrm{DM}$. The dependence of $M_\mathrm{PNS}$ and thus the g-mode frequency on both $M_\mathrm{DM}$ and $\beta_\mathrm{i,b}$ complements the relations found in Section~\ref{subsec:DM_signal}, though the detailed frequency information awaits post-bounce neutrino transport simulations. A joint analysis of the bounce and PNS g-model GW signals can unveil the presence of DM if admixed inside AICs. 

\begin{figure}[t!]
	\plotone{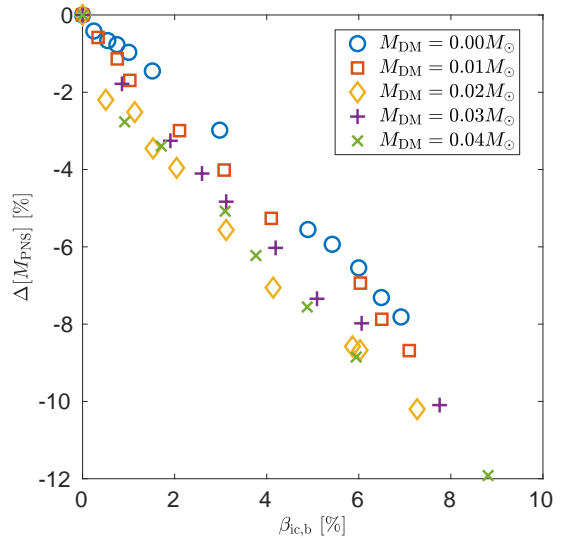}
	\caption{Same as Fig.~\ref{fig:tb_rhoc}, but for the relative differences of $M_\mathrm{PNS}$ defined in Section~\ref{subsec:mpns}.\label{fig:dmpns}}. 
\end{figure}

\section{Conclusion} \label{sec:conclu}
We have performed axisymmetric hydrodynamical simulations of AIC of DM admixed WDs, with uniform rotation initially. With DM admixture, the early contraction is accelerated by the deeper central gravitational potential well. However, the decrement of the total WD mass due to admixed DM eventually delays the plunge and bounce phase from $\sim30~\mathrm{ms}$ for $M_\mathrm{DM}=0$ to $\sim100~\mathrm{ms}$ for $M_\mathrm{DM}=0.04~M_\odot$. Also, the central density at bounce and equilibrium are smaller for a larger $M_\mathrm{DM}$. Key characteristics for the collapse and bounce of AIC, $\mathcal{O}$, such as the bounce time and central density, generally depend on $M_\mathrm{DM}$ and $\beta_\mathrm{ic,b}$, in a factorized form: $\mathcal{O} = f(\beta_\mathrm{ic,b})\mathcal{O}^{(0)}(M_\mathrm{DM})$, where $\mathcal{O}^{(0)}$ is for the non-rotating model with possible DM admixture.

The above results determine the dependence of GW signals during collapse and early post-bounce phases on $\beta_\mathrm{ic,b}$ and $M_\mathrm{DM}$. With the same $\Omega_\mathrm{ini}$ and initial $\rho_\mathrm{c}$, GW is enhanced by admixing DM. For models reaching the same $\beta_\mathrm{ic,b}$, the GW amplitude is decreased by the DM admixture due to the less compact inner core. The ratios between GW amplitudes of the three dominant spikes around bounce time show a strong dependence on $M_\mathrm{DM}$ and can be used as indicators of DM admixing. Assuming that the microphysics inputs can be well constrained in the future, $M_\mathrm{DM}$ and $\beta_\mathrm{ic,b}$ can both be retrieved from the observed GW signals. On the other hand, if the DM core mass $M_\mathrm{DM} \ge 0.03~M_\odot$, its existence can still be inferred despite the uncertainties of the nuclear matter EOS.

Although our simulations are Newtonian with GR modification of the gravitational potential, we believe that our conclusion on DM effects would not be changed significantly in full GR modeling since the bounce signal matches quite well to GR simulations with the CFC approximation. The parameterization scheme for electron capture during the collapse phase \citep{2005ApJ...633.1042L} should not affect our current conclusion qualitatively. However, to investigate the dependence of the PNS g-mode frequency, as well as the explosion energy and ejected mass (and thus light curves) on DM mass, the long-term post-bounce neutrino transport is indispensable. We leave the neutrino-transport simulation and the DM effects on electromagnetic and neutrino signals for a future work. 

\acknowledgments
\textit{Acknowledgments --- }We thank Evan O'Connor for making his neutrino transport code \texttt{GR1D} open-source and \url{stellarcollapse.org} for sharing tools of supernova equations of state and several gravitational wave catalogs.  We also thank Ken'ichi Nomoto, Bernhard M\"uller and Ronald Taam for stimulating discussions. This work is partially supported by a grant from the Research Grant Council of Hong Kong (Project 14300317). S.C.L. is supported by World Premier International Research Center Initiative (WPI), MEXT, Japan and JSPS KAKENHI Grant Numbers JP26400222, JP16H02168, JP17K05382. We acknowledge the support of the CUHK Central High Performance Computing Cluster, on which the computation in this work has been performed.	

\appendix
\section{GR1D simulation} \label{app:GR1D}
A parametrized electron capture scheme is used in the two-dimensional simulations, and the $\rho-Y_e$ profile parametrization is obtained from \texttt{GR1D} \citep{2015ApJS..219...24O} simulations. \texttt{GR1D} solves the neutrino transport problem with a two-moment method with an analytic closure (the so-called M1 scheme), and the most important neutrino emission, absorption, and scattering reactions \deleted{(Table~\ref{tab:nulib})} are included with the rates provided by \texttt{NuLib} \citep[\added{Table 1 in }][]{2015ApJS..219...24O}. The parameters used for this neutrino transport simulation are similar to those used in a $15~M_\odot$ core-collapse simulation presented in the code paper. Fig.~\ref{fig:gr1d} shows the results of \texttt{GR1D} simulations with the \texttt{LS220} EOS, including central density evolution, post-bounce shock propagation, and luminosity and root mean squared (rms) energy of neutrinos. These results are consistent with those in the literature \citep{2006ApJ...644.1063D,2006A&A...450..345K}.


\begin{figure*}[h!]
	\setcounter{figure}{0}
	\renewcommand{\thefigure}{A\arabic{figure}}
	\plotone{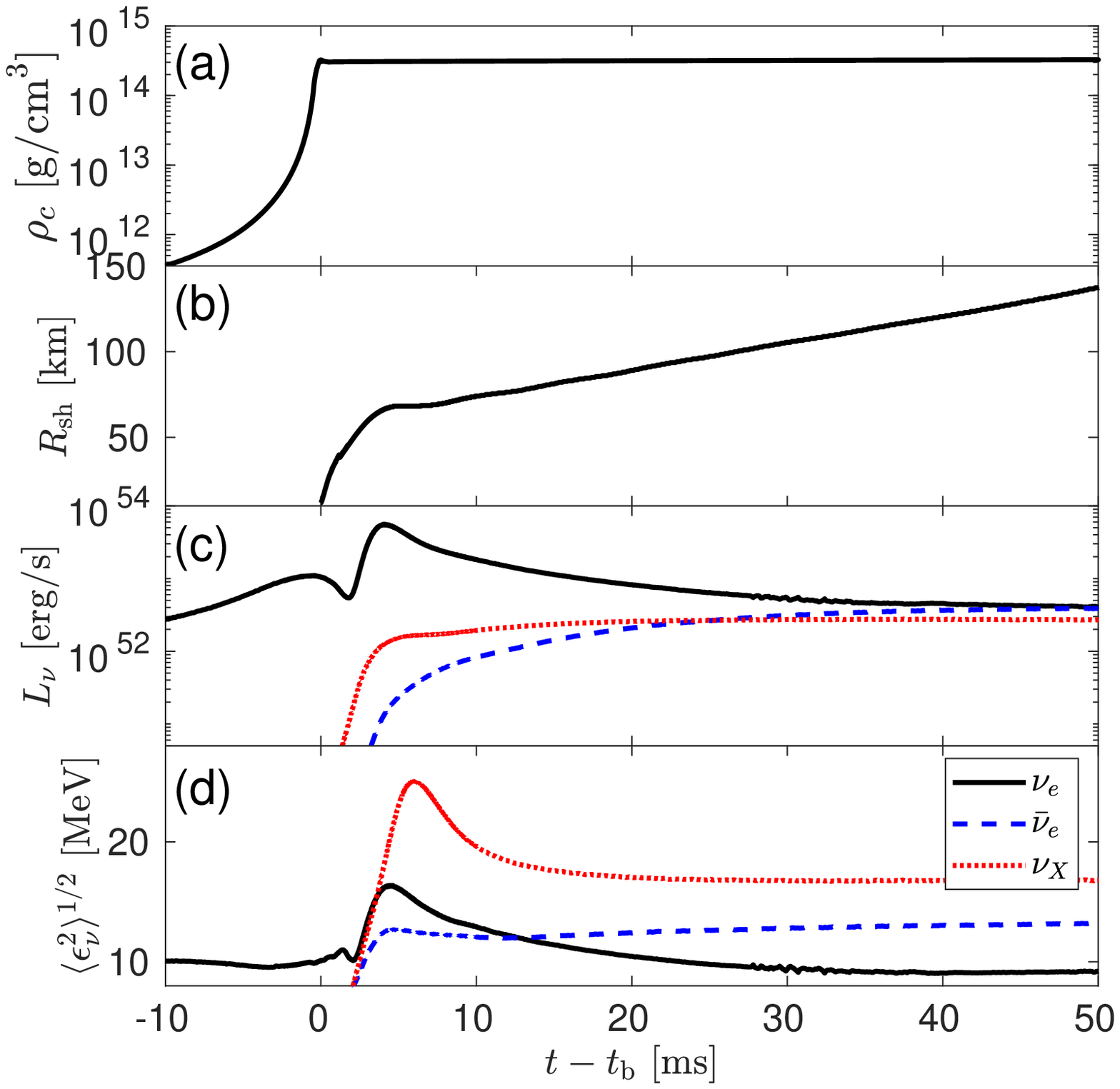}
	\caption{Time evolution of a) central density, b) post-bounce shock propagation, c) neutrino luminosities, and d) rms energy of neutrinos from a spherically symmetric AIC simulation using the \texttt{GR1D} code, with the \texttt{LS220} EOS. \label{fig:gr1d} }
\end{figure*}

\section{EOS dependence} \label{app:EOS}
In this appendix, we study the effects of different EOS models on rotating AIC simulations without DM admixture, with the parametrized profiles ($\rho-Y_e$) from \texttt{GR1D} simulations with different EOSs. The initial WD has a $\Omega_\mathrm{ini}=5~\mathrm{rad/s}$. The chosen EOSs are the widely used ones, \texttt{HShen} \citep{2011ApJS..197...20S}, \texttt{LS220} \citep{1991NuPhA.535..331L}, and \texttt{SFHo} \citep{2013ApJ...774...17S}, which represent high, moderate, and low stiffness. They all satisfy the constraint imposed by the precise measurement of maximum pulsar mass (Pulsar J0348+0432, $2.01\pm0.04~M_\odot$, \cite{2013Sci...340..448A}). Note that although \texttt{SFHo} is declared to be the most consistent to all constraints for nuclear matter EOSs, the reason is that the relativistic mean field (RMF) parameters come from optimizing the fitting likelihood to the neutron star mass-radius curve. So for our major goal of investigating DM effects, we choose the most explored \texttt{LS220} EOS as the standard. The resulting GWs and their spectra are presented in Fig.~\ref{fig:gw_eos}. The waveforms are almost identical around the time of core bounce, except for tiny differences in the GW amplitude ($h_{2,3}$), while the post-bounce ring-down signals show quasi-periodic cycles with different periods. It is further found that if the GW frequency is normalized with the dynamical frequency
\begin{equation} \label{eq:fdyn}
f_\mathrm{dyn} = \sqrt{G\rho_\mathrm{c}},
\end{equation}
the differences in peak frequencies of the prominent GW modes disappear, and the different EOSs give very similar spectra with slightly different amplitudes.
\begin{figure}[t!]
	\setcounter{figure}{0}
	\renewcommand{\thefigure}{B\arabic{figure}}
	\plottwo{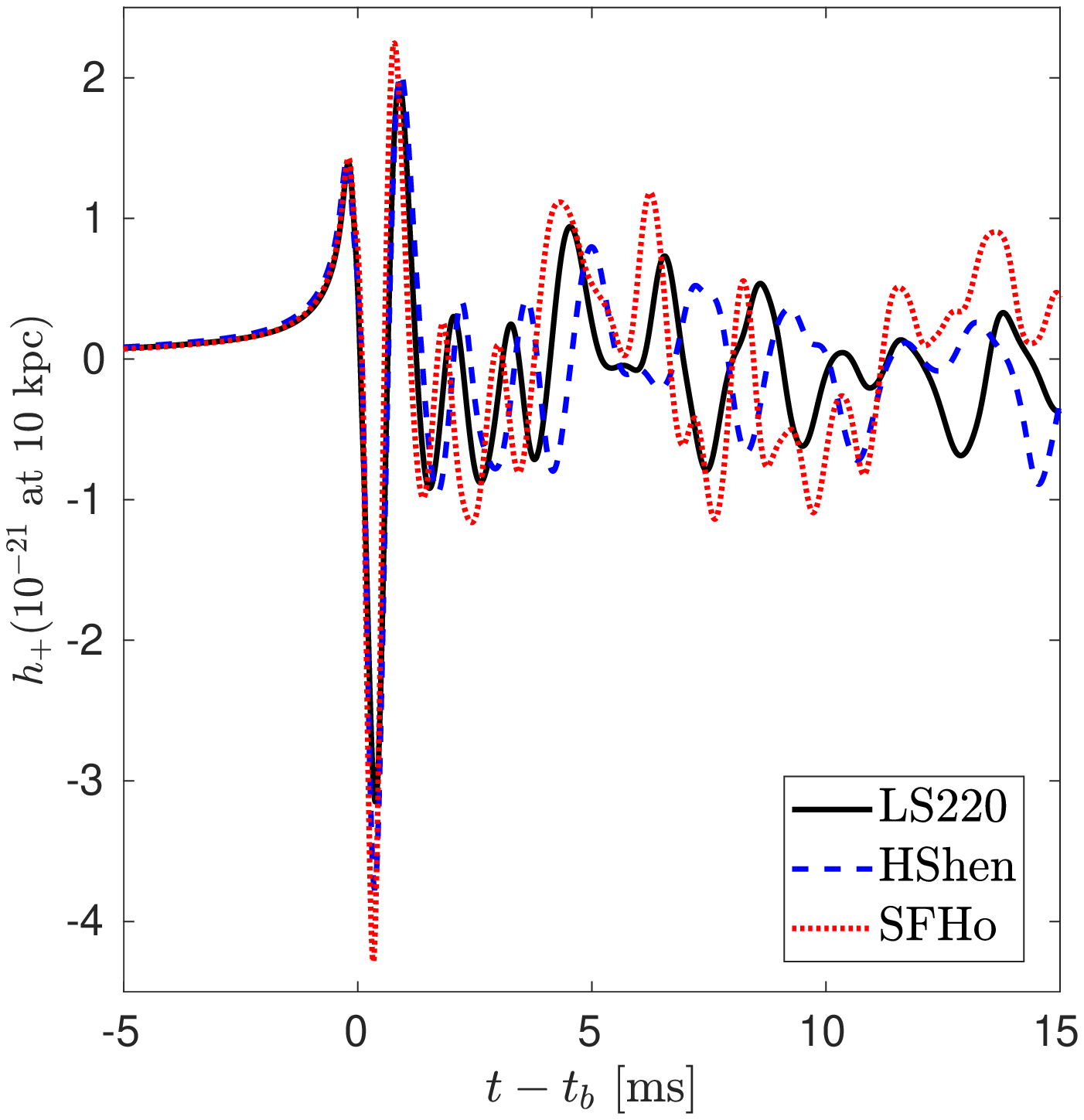}{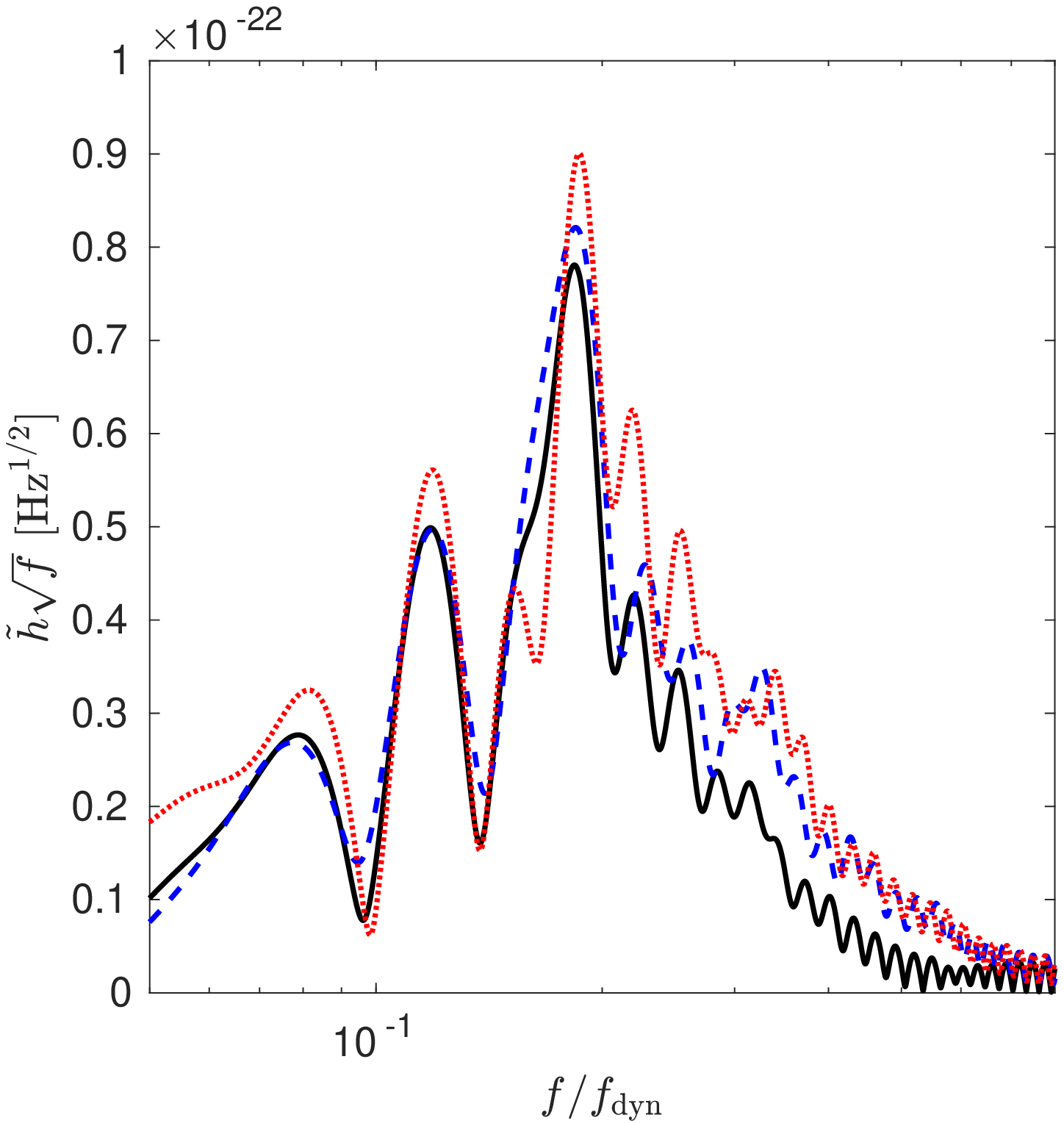}
	\caption{GW waveforms (left) and their Fourier spectra (right) from the AIC of a rotating WD ($\Omega_\mathrm{ini}=5~\mathrm{rad/s}$) with three different nuclear matter EOSs. In the spectra, the frequency has been normalized to the dynamical frequency $f_\mathrm{dyn}$ defined in Eq. \ref{eq:fdyn}. \label{fig:gw_eos}}
\end{figure}

\section{Central density of the progenitor WD} \label{app:rho_ci}
As discussed in Section~\ref{subsec:init}, the initial central density $\rho_\mathrm{c}$ of an AIC progenitor has not been accurately determined from stellar evolution calculations yet. Three different initial $\rho_\mathrm{c}$ ($5.0,~2.5,~1.0\times10^{10}~\mathrm{g/cm^3}$) have been used to test the variations in bounce GW signal, with $\Omega_\mathrm{ini}=5,~5,~4~\mathrm{rad/s}$ and $\beta_\mathrm{ic,b}= 1.52,~3.7,~7.2\%$. Hydrodynamical simulations are performed with the same settings such as \texttt{LS220} EOS and the $Y_e(\rho)$ relation. Their waveforms with rescaled GW amplitudes to match $h_1$ are shown in Fig.~\ref{fig:gw_rhoc}. The excellent match of the first three spikes suggests that the variations in the ratios $h_2/h_1$ and $h_3/h_1$ for different initial $\rho_\mathrm{c}$ are very small. Therefore, the usage of these ratios for identifying DM admixture is not affected by the uncertainty in $\rho_\mathrm{c}$ of the AIC progenitor.
\begin{figure}
	\centering
	\setcounter{figure}{0}
	\renewcommand{\thefigure}{C\arabic{figure}}	
	\includegraphics[width=0.5\textwidth]{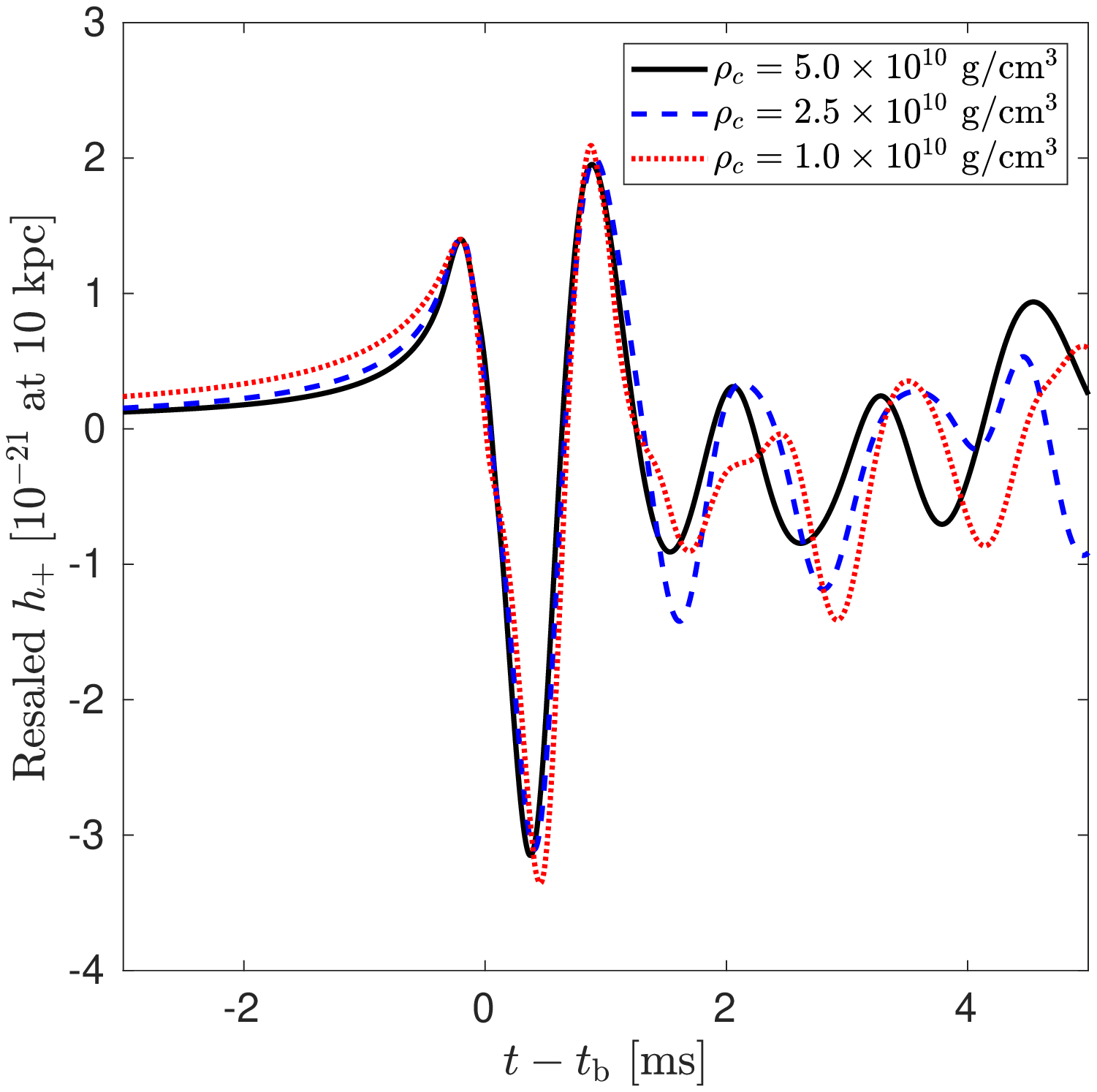}	
	\caption{GW waveforms from the AICs of rotating WDs with three different initial $\rho_\mathrm{c}$ ($\Omega_\mathrm{ini}$), $5.0\times10^{10}~\mathrm{g/cm^3}$~($5~\mathrm{rad/s}$), $2.5\times10^{10}~\mathrm{g/cm^3}$~($5~\mathrm{rad/s}$), $1.0\times10^{10}~\mathrm{g/cm^3}$~($4~\mathrm{rad/s}$), all with \texttt{LS220} EOS. The GW amplitudes are multiplied by a constant to match the peak amplitudes before bounce ($h_1$). \label{fig:gw_rhoc}}
\end{figure}

\section{Convergence test \label{app:converge}}
The convergence of GW waveform for the same model, but simulated with different resolutions, has been an issue for the CCSN community \citep{2009CQGra..26f3001O}. In a non-rotating model, the asphericity comes from convective motions seeded by stochastic perturbations, eg. grid noise, and so a precise match of GW waveform for different resolutions is not expected. Nonetheless, in our non-rotating and DM-absent AIC model, the GW emission from convection emerges after 10~ms postbounce and the maximum $h_+$ reaches $\sim0.1\times10^{-21}$ at 10~kpc. This is a factor of 5 smaller than the slowest rotating model in this paper. To check the convergence for the rotating models, we performed two additional simulations with higher resolutions for the moderately rotating AIC model (R5-DM0, $\rho_{\rm c,ini}=5\times10^{10}~{\rm g/cm^3}$ and $\Omega_{\rm ini}=5~{\rm rad/s}$). The resulting GW waveforms are compared with that of the fiducial run in Fig.~\ref{fig:gw_converge}. Remarkably, the waveforms match excellently for $t\lesssim t_{\rm b}+5~{\rm ms}$, while the later oscillations differ slightly due to the contribution from convection.

\begin{figure}
	\centering
	\setcounter{figure}{0}
	\renewcommand{\thefigure}{D\arabic{figure}}	
	\includegraphics[width=0.5\textwidth]{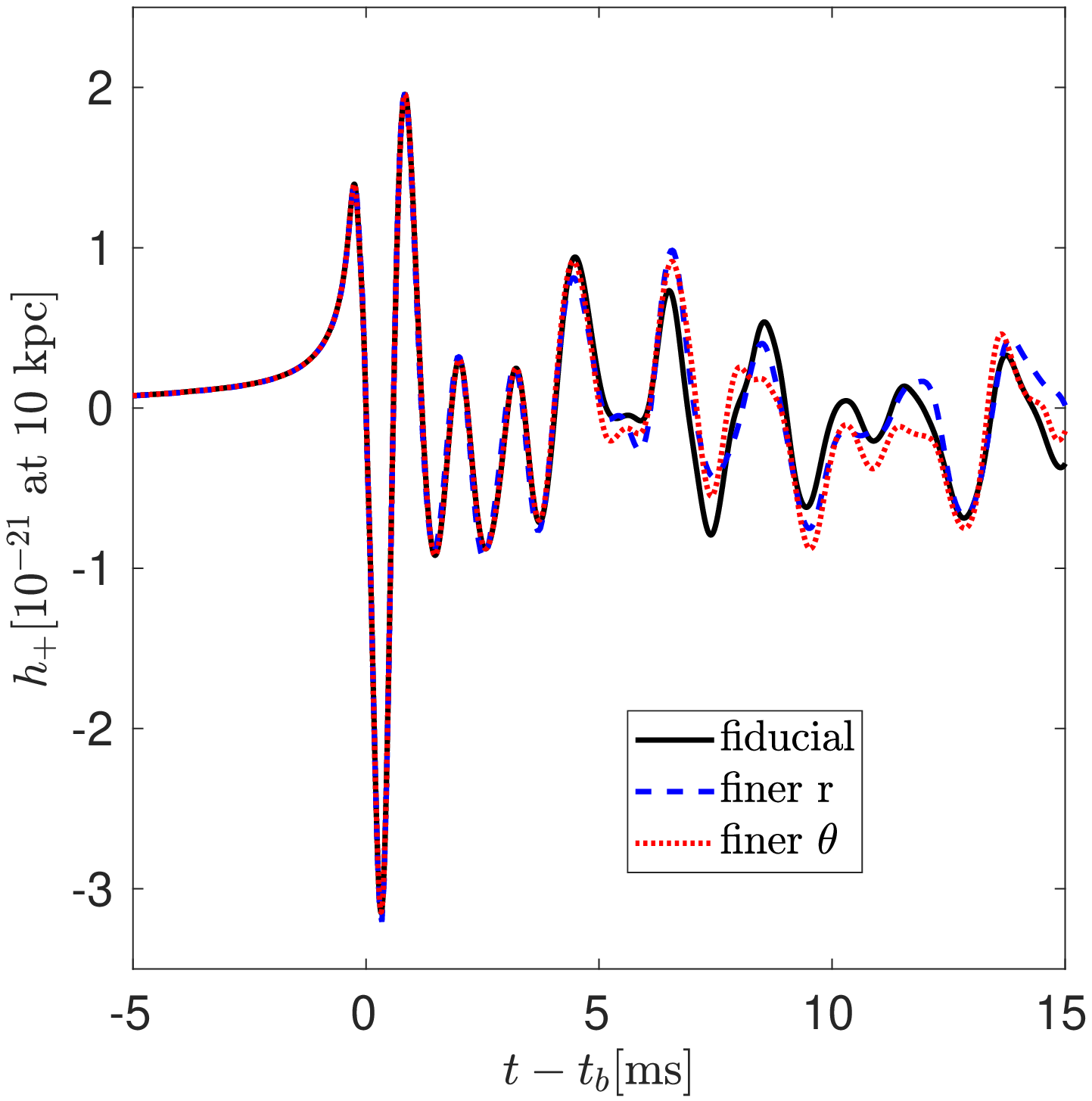}	
	\caption{Convergence test for GW waveform of the R5-DM0 model, with different resolutions. The fiducial run is with $2^\circ$ for $\theta$ grid and $370$~m for $r$ grid near the center. Finer $r$ is with the same $\theta$ grid as the fiducial run and $220$~m for the $r$ grid near the center. Finer $\theta$ is with $1.5^\circ$ for the $\theta$ grid and the same $r$ grid as the fiducial run. \label{fig:gw_converge}}
\end{figure}

\clearpage
\added{
\section{Extended table for the results of all the models \label{app:extend}}
\begin{table}[th!]
	\setcounter{table}{0}
	\renewcommand{\thetable}{E\arabic{table}}
	\centering
	\caption{\added{Extended table for the results of the models with a specified initial angular velocity $\Omega_{\rm ini}$ and admixed DM mass $M_{\rm DM}$.  $M_\mathrm{NM}$ is the NM mass. $J$ is the total initial angular momentum. $\beta_\mathrm{ini}$ and $\beta_\mathrm{ic,b}$ are the initial ratio of rotational energy to gravitational energy and that of the inner core at the bounce. $R_\mathrm{e}$ and $R_\mathrm{p}$ are the equatorial and polar radii of the white dwarfs, respectively. $t_{\rm b}$, $\rho_{\rm c,b}$ and $M_{\rm ic,b}$ are the time, central density and mass of the inner core at the bounce. $M_{\rm PNS}$ is the PNS mass at 50 ms after the bounce.} \label{tab:all_models}}
	\begin{tabular}{ccccccccccccc}
		\toprule
		\multirow{2}{*}{Model} &
			$\Omega_{\rm ini}$ &
			$M_\mathrm{DM}$ &
			$M_\mathrm{NM}$ &
			$J$ & 
			$\beta_\mathrm{ini}$ &
			$R_\mathrm{e}$ &
			$R_\mathrm{p}/R_\mathrm{e}$ &
			$\beta_\mathrm{ic,b}$ &
			$t_{\rm b}$ &
			$\rho_{\rm c,b}$ &
			$M_{\rm ic,b}$ &
			$M_{\rm PNS}$ \\
			&
			$[\rm rad/s]$ &
			$[M_\odot]$ &
			$[M_\odot]$ & 
			$[10^{50}\mathrm{erg\cdot s}]$ & 
			$[\mathrm{\%}]$ &
			$[\mathrm{km}]$ &
			  &
			$[\mathrm{\%}]$ &
			[ms] &
			$[10^{14}\rm g/cm^3]$ &
			$[M_\odot]$ &
			$[M_\odot]$ \\ \hline
		R0-DM0  & 0.0  &0      & 1.448 &  --   &   --   &  816  &  1 & -- & 32.8 & 3.99 & 0.56 & 1.26  \\
		R0-DM1   & 0.0  &0.01 & 1.402 &  --   &   --   &  888  &  1 & -- & 37.7 & 3.90 & 0.54 & 1.21  \\
		R0-DM2  & 0.0  &0.02 & 1.355 &  --   &   --   &  977  &  1 & -- & 45.2 & 3.80 & 0.53 & 1.15  \\
		R0-DM3  & 0.0  &0.03 & 1.304 &  --   &   --   & 1113  &  1 & -- & 58.6 & 3.69 & 0.51 & 1.07 \\
		R0-DM4  & 0.0  &0.04 & 1.249 &  --   &   --   & 1313  &  1 & -- & 86.1 & 3.55 & 0.49 & 0.99 \\ \hline
		R2-DM0  & 2.0  &0      & 1.450 &  0.04   &   0.04   &  820  &  0.995 & 0.25 & 32.9 & 3.98 & 0.56 & 1.26  \\
		R2-DM1   & 2.0  &0.01 & 1.405 &  0.04   &   0.05   &  893  &  0.990 & 0.34 & 37.8 & 3.88 & 0.55 & 1.20  \\
		R2-DM2  & 2.0  &0.02 & 1.358 &  0.05   &   0.07   &  987  &  0.990 & 0.50 & 45.4 & 3.77 & 0.54 & 1.13  \\
		R2-DM3  & 2.0  &0.03 & 1.309 &  0.06   &   0.11   & 1130  &  0.980 & 0.86 & 59.0 & 3.63 & 0.53 & 1.05 \\
		R2-DM4  & 2.0  &0.04 & 1.257 &  0.08   &   0.19   & 1347  &  0.970 & 1.71 & 87.3 & 3.44 & 0.51 & 0.96 \\ \hline
		R3-DM0  & 3.0  &0      & 1.451 &  0.05   &   0.09   &  828  &  0.985 & 0.55 & 33.0 & 3.94 & 0.56 & 1.26  \\
		R3-DM1   & 3.0  &0.01 & 1.407 &  0.06   &   0.12   &  902  &  0.980 & 0.75 & 38.0 & 3.83 & 0.55 & 1.20  \\
		R3-DM2  & 3.0  &0.02 & 1.361 &  0.07   &   0.16   &  1002  &  0.970 & 1.13 & 45.6 & 3.71 & 0.55 & 1.12  \\
		R3-DM3  & 3.0  &0.03 & 1.314 &  0.09   &   0.25   & 1159  &  0.956 & 1.91 & 59.3 & 3.52 & 0.54 & 1.03 \\
		R3-DM4  & 3.0  &0.04 & 1.266 &  0.13   &   0.43   & 1395  &  0.932 & 3.77 & 88.4 & 3.26 & 0.53 & 0.93 \\ \hline
		R3.5-DM0  & 3.5  &0      & 1.453 &  0.06   &   0.12   &  832  &  0.980 & 0.75 & 33.0 & 3.92 & 0.57 & 1.25  \\
		R3.5-DM1   & 3.5  &0.01 & 1.409 &  0.07   &   0.16   &  902  &  0.975 & 1.03 & 38.1 & 3.81 & 0.56 & 1.19  \\
		R3.5-DM2  & 3.5  &0.02 & 1.364 &  0.09   &   0.22   &  1007  &  0.966 & 1.53 & 45.8 & 3.67 & 0.56 & 1.11  \\
		R3.5-DM3  & 3.5  &0.03 & 1.318 &  0.11   &   0.34   & 1165  &  0.946 & 2.59 & 59.7 & 3.47 & 0.55 & 1.02 \\
		R3.5-DM4  & 3.5  &0.04 & 1.273 &  0.16   &   0.60   & 1430  &  0.900 & 4.88 & 89.4 & 3.13 & 0.55 & 0.91 \\ \hline
		R5-DM0  & 5.0  &0      & 1.458 &  0.09   &    0.25   &  849  &  0.956 & 1.52 & 33.2 & 3.85 & 0.59 & 1.25  \\
		R5-DM1   & 5.0  &0.01 & 1.416 &  0.11   &    0.33   &  934  &  0.942 & 2.11 & 38.4 & 3.73 & 0.59 & 1.17 \\
		R5-DM2  & 5.0 &0.02 & 1.374 &  0.13   &    0.46   & 1043  &  0.928 & 3.12 & 46.3 & 3.55 & 0.58 & 1.09 \\
		R5-DM3  & 5.0  &0.03 & 1.333 &  0.16   &    0.71   & 1243  &  0.896 & 5.09 & 60.9 & 3.23 & 0.59 & 0.99 \\
		R5-DM4 \footnote{Also Rmax-DM4.}  & 5.0  &0.04 & 1.303 &  0.25   &    1.33   & 1809  &  0.726 & 8.81 & 93.1 & 2.73 & 0.59 & 0.87 \\ \hline
		R7-DM0  & 7.0  &0      & 1.469 &  0.13   &    0.50   &  875  &  0.918 & 2.98 & 33.6 & 3.74 & 0.63 & 1.23  \\
		R7-DM1   & 7.0  &0.01 & 1.430 &  0.15   &    0.66   &  982  &  0.900 & 4.10 & 39.0 & 3.55 & 0.63 & 1.15 \\
		R7-DM2  & 7.0 &0.02 & 1.394 &  0.19   &    0.94   & 1147  &  0.852 & 5.86 & 47.3 & 3.27 & 0.63 & 1.05 \\ \hline
		R9-DM0  & 9.0  &0      & 1.485 &  0.18   &    0.86   &  939  &  0.869 & 4.89 & 34.2 & 3.57 & 0.66 & 1.19  \\
		R9-DM1   & 9.0  &0.01 & 1.451 &  0.21   &    1.15   &  1113  &  0.794 & 6.50 & 39.8 & 3.31 & 0.67 & 1.11 \\ \hline
		Rmax-DM0 & 10.9 & 0     & 1.506 & 0.23   &   1.32	  &  1165  &  0.673 & 6.92 & 34.9 & 3.37 & 0.70 & 1.17 \\
		Rmax-DM1 & 9.5 & 0.01  & 1.459 & 0.23   &   1.30	  &  1256  &  0.680 & 7.10 & 40.1 & 3.24 & 0.68 & 1.10 \\
		 Rmax-DM2 & 8.0 & 0.02  & 1.409 & 0.23   &   1.27	  &  1333  &  0.708 & 7.27 & 48.0 & 3.10 & 0.66 & 1.03 \\
		 Rmax-DM3 & 6.50 & 0.03  & 1.358 & 0.23   &   1.28	  &  1519  &  0.704 & 7.76 & 62.5 & 2.90 & 0.63 & 0.96 \\ \hline
		 
	\end{tabular}
\end{table}
\clearpage
\begin{table}[t!]
	\renewcommand{\thetable}{E\arabic{table}}
	\centering
	\caption{\added{Same as Table~\ref{tab:all_models} but for the models with a specified $\beta_\mathrm{ic,b}$ and admixed DM mass $M_{\rm DM}$. Some models are the same as those in Table~\ref{tab:all_models}.} \label{tab:beta_seris}}
	\begin{tabular}{ccccccccccccc}
		\toprule
		\multirow{2}{*}{Model} &
		$\Omega_{\rm ini}$ &
		$M_\mathrm{DM}$ &
		$M_\mathrm{NM}$ &
		$J$ & 
		$\beta_\mathrm{ini}$ &
		$R_\mathrm{e}$ &
		$R_\mathrm{p}/R_\mathrm{e}$ &
		$\beta_\mathrm{ic,b}$ &
		$t_{\rm b}$ &
		$\rho_{\rm c,b}$ &
		$M_{\rm ic,b}$ &
		$M_{\rm PNS}$ \\
		&
		$[\rm rad/s]$ &
		$[M_\odot]$ &
		$[M_\odot]$ & 
		$[10^{50}\mathrm{erg\cdot s}]$ & 
		$[\mathrm{\%}]$ &
		$[\mathrm{km}]$ &
		&
		$[\mathrm{\%}]$ &
		[ms] &
		$[10^{14}\rm g/cm^3]$ &
		$[M_\odot]$ &
		$[M_\odot]$ \\ \hline
		$\beta1$-DM0  & 4.1  &0      & 1.455 &  0.07   &   0.17   &  832  &  0.975 & 1.00 & 33.1 & 3.90 & 0.58 & 1.25  \\
		$\beta1$-DM1\footnote{Also the model R3.5-DM1.}   & 3.5  &0.01 & 1.409 &  0.07   &   0.16   &  902  &  0.975 & 1.03 & 38.1 & 3.81 & 0.56 & 1.19  \\
		$\beta1$-DM2\footnote{Also the model R3-DM2.}  & 3.0  &0.02 & 1.361 &  0.07   &   0.16   &  1002  &  0.970 & 1.13 & 45.6 & 3.71 & 0.55 & 1.12  \\
		$\beta1$-DM3  & 2.2  &0.03 & 1.309 &  0.07   &   0.12   & 1130  &  0.980 & 0.99 & 59.1 & 3.61 & 0.53 & 1.04 \\
		$\beta1$-DM4  & 1.5  &0.04 & 1.253 &  0.06   &   0.10   & 1333  &  0.985 & 0.91 & 86.8 & 3.50 & 0.50 & 0.96 \\ \hline
		$\beta3$-DM0\footnote{Also the model R7-DM0.}  & 7.0  &0      & 1.469 &  0.13   &    0.50   &  875  &  0.918 & 2.98 & 33.6 & 3.74 & 0.63 & 1.23   \\
		$\beta3$-DM1   & 6.1  &0.01 & 1.423 &  0.13   &   0.49   &  948  &  0.918 & 3.08 & 38.7 & 3.64 & 0.61 & 1.16  \\
		$\beta3$-DM2\footnote{Also the model R5-DM2.}  & 5.0 &0.02 & 1.374 &  0.13   &    0.46   & 1043  &  0.928 & 3.12 & 46.3 & 3.51 & 0.58 & 1.09  \\
		$\beta3$-DM3  & 3.8  &0.03 & 1.321 &  0.12   &   0.41   & 1176  &  0.932 & 3.12 & 60.0 & 3.43 & 0.56 & 1.01 \\
		$\beta3$-DM4  & 1.3  &0.04 & 1.263 &  0.12   &   0.35   & 1374  &  0.946 & 3.10 & 88.1 & 3.33 & 0.53 & 0.94 \\ \hline
		$\beta6$-DM0  & 10.0  &0      & 1.495 &  0.21   &    1.09   &  997  &  0.790 & 6.00 & 34.6 & 3.46 & 0.68 & 1.18   \\
		$\beta6$-DM1   & 8.6  &0.01 & 1.447 &  0.20   &   1.04   &  1070  &  0.802 & 6.03 & 39.7 & 3.36 & 0.66 & 1.13  \\
		$\beta6$-DM2  & 7.1 &0.02 & 1.396 &  0.19   &    0.97   & 1159  &  0.818 & 6.03 & 47.5 & 3.26 & 0.63 & 1.05  \\
		$\beta6$-DM3  & 5.6  &0.03 & 1.341 &  0.19   &   0.90   & 1294  &  0.839 & 6.06 & 61.5 & 3.11 & 0.60 & 0.98 \\
		$\beta6$-DM4  & 4.0  &0.04 & 1.281 &  0.18   &   0.79   & 1481  &  0.865 & 5.94 & 90.4 & 2.99 & 0.56 & 0.90 \\ \hline
	\end{tabular}
\end{table}
}

\bibliographystyle{aasjournal}
\bibliography{mybib}

\begin{thebibliography}{}
\expandafter\ifx\csname natexlab\endcsname\relax\def\natexlab#1{#1}\fi
\providecommand{\url}[1]{\href{#1}{#1}}
\providecommand{\dodoi}[1]{doi:~\href{http://doi.org/#1}{\nolinkurl{#1}}}
\providecommand{\doeprint}[1]{\href{http://ascl.net/#1}{\nolinkurl{http://ascl.net/#1}}}
\providecommand{\doarXiv}[1]{\href{https://arxiv.org/abs/#1}{\nolinkurl{https://arxiv.org/abs/#1}}}

\bibitem[{{Abdikamalov} {et~al.}(2010){Abdikamalov}, {Ott}, {Rezzolla},
  {Dessart}, {Dimmelmeier}, {Marek}, \& {Janka}}]{2010PhRvD..81d4012A}
{Abdikamalov}, E.~B., {Ott}, C.~D., {Rezzolla}, L., {et~al.} 2010, \prd, 81,
  044012

\bibitem[{{Andresen} {et~al.}(2019){Andresen}, {M{\"u}ller}, {Janka}, {Summa},
  {Gill}, \& {Zanolin}}]{2019MNRAS.486.2238A}
{Andresen}, H., {M{\"u}ller}, E., {Janka}, H.~T., {et~al.} 2019, \mnras, 486,
  2238

\bibitem[{{Antoniadis} {et~al.}(2013){Antoniadis}, {Freire}, {Wex}, {Tauris},
  {Lynch}, {van Kerkwijk}, {Kramer}, {Bassa}, {Dhillon}, {Driebe}, {Hessels},
  {Kaspi}, {Kondratiev}, {Langer}, {Marsh}, {McLaughlin}, {Pennucci}, {Ransom},
  {Stairs}, {van Leeuwen}, {Verbiest}, \& {Whelan}}]{2013Sci...340..448A}
{Antoniadis}, J., {Freire}, P.~C.~C., {Wex}, N., {et~al.} 2013, Science, 340,
  448

\bibitem[{{Baiotti} {et~al.}(2007){Baiotti}, {de Pietri}, {Manca}, \&
  {Rezzolla}}]{2007PhRvD..75d4023B}
{Baiotti}, L., {de Pietri}, R., {Manca}, G.~M., \& {Rezzolla}, L. 2007, \prd,
  75, 044023

\bibitem[{{Barkana}(2018)}]{2018Natur.555...71B}
{Barkana}, R. 2018, \nat, 555, 71

\bibitem[{{Barsotti} {et~al.}(2018){Barsotti}, {Fritschel}, {Evans}, \&
  {Gras}}]{LIGO_v5}
{Barsotti}, L., {Fritschel}, P., {Evans}, M., \& {Gras}, S. 2018, Tech. Rep.
  {LIGO-T1800042-v5}.
\newblock \url{dcc.ligo.org/LIGO-T1800042/public}

\bibitem[{{Bertone} \& {Hooper}(2018)}]{2018RvMP...90d5002B}
{Bertone}, G., \& {Hooper}, D. 2018, RvMP, 90, 045002

\bibitem[{{Bowman} {et~al.}(2018){Bowman}, {Rogers}, {Monsalve}, {Mozdzen}, \&
  {Mahesh}}]{2018Natur.555...67B}
{Bowman}, J.~D., {Rogers}, A. E.~E., {Monsalve}, R.~A., {Mozdzen}, T.~J., \&
  {Mahesh}, N. 2018, \nat, 555, 67

\bibitem[{{Bramante}(2015)}]{2015PhRvL.115n1301B}
{Bramante}, J. 2015, \prl, 115, 141301

\bibitem[{{Brito} {et~al.}(2015){Brito}, {Cardoso}, \&
  {Okawa}}]{2015PhRvL.115k1301B}
{Brito}, R., {Cardoso}, V., \& {Okawa}, H. 2015, \prl, 115, 111301

\bibitem[{{Cerd{\'a}-Dur{\'a}n} {et~al.}(2013){Cerd{\'a}-Dur{\'a}n}, {DeBrye},
  {Aloy}, {Font}, \& {Obergaulinger}}]{2013ApJ...779L..18C}
{Cerd{\'a}-Dur{\'a}n}, P., {DeBrye}, N., {Aloy}, M.~A., {Font}, J.~A., \&
  {Obergaulinger}, M. 2013, \apj, 779, L18

\bibitem[{{Cerd{\'a}-Dur{\'a}n} {et~al.}(2007){Cerd{\'a}-Dur{\'a}n}, {Quilis},
  \& {Font}}]{2007CoPhC.177..288C}
{Cerd{\'a}-Dur{\'a}n}, P., {Quilis}, V., \& {Font}, J.~A. 2007, CoPhC, 177, 288

\bibitem[{{Cerme{\~n}o} {et~al.}(2017){Cerme{\~n}o}, {P{\'e}rez-Garc{\'{\i}}a},
  \& {Silk}}]{2017PASA...34...43C}
{Cerme{\~n}o}, M., {P{\'e}rez-Garc{\'{\i}}a}, M.~{\'A}., \& {Silk}, J. 2017,
  \pasa, 34, e043

\bibitem[{{Choplin} {et~al.}(2017){Choplin}, {Coc}, {Meynet}, {Olive}, {Uzan},
  \& {Vangioni}}]{2017A&A...605A.106C}
{Choplin}, A., {Coc}, A., {Meynet}, G., {et~al.} 2017, \aap, 605, A106

\bibitem[{{Dessart} {et~al.}(2006){Dessart}, {Burrows}, {Ott}, {Livne}, {Yoon},
  \& {Langer}}]{2006ApJ...644.1063D}
{Dessart}, L., {Burrows}, A., {Ott}, C.~D., {et~al.} 2006, \apj, 644, 1063

\bibitem[{{Dimmelmeier} {et~al.}(2002){Dimmelmeier}, {Font}, \&
  {M{\"u}ller}}]{2002A&A...393..523D}
{Dimmelmeier}, H., {Font}, J.~A., \& {M{\"u}ller}, E. 2002, \aap, 393, 523

\bibitem[{{Dimmelmeier} {et~al.}(2008){Dimmelmeier}, {Ott}, {Marek}, \&
  {Janka}}]{2008PhRvD..78f4056D}
{Dimmelmeier}, H., {Ott}, C.~D., {Marek}, A., \& {Janka}, H.-T. 2008, \prd, 78,
  064056

\bibitem[{{Einasto} {et~al.}(1974){Einasto}, {Saar}, {Kaasik}, \&
  {Chernin}}]{1974Natur.252..111E}
{Einasto}, J., {Saar}, E., {Kaasik}, A., \& {Chernin}, A.~D. 1974, \nat, 252,
  111

\bibitem[{{Ellis} {et~al.}(2018){Ellis}, {H{\"u}tsi}, {Kannike}, {Marzola},
  {Raidal}, \& {Vaskonen}}]{2018PhRvD..97l3007E}
{Ellis}, J., {H{\"u}tsi}, G., {Kannike}, K., {et~al.} 2018, \prd, 97, 123007

\bibitem[{{Farrow} {et~al.}(2019){Farrow}, {Zhu}, \&
  {Thrane}}]{2019ApJ...876...18F}
{Farrow}, N., {Zhu}, X.-J., \& {Thrane}, E. 2019, \apj, 876, 18

\bibitem[{{Fink} {et~al.}(2018){Fink}, {Kromer}, {Hillebrandt}, {R{\"o}pke},
  {Pakmor}, {Seitenzahl}, \& {Sim}}]{2018A&A...618A.124F}
{Fink}, M., {Kromer}, M., {Hillebrandt}, W., {et~al.} 2018, \aap, 618, A124

\bibitem[{{Finn} \& {Evans}(1990)}]{1990ApJ...351..588F}
{Finn}, L.~S., \& {Evans}, C.~R. 1990, \apj, 351, 588

\bibitem[{{Fryer} {et~al.}(1999){Fryer}, {Benz}, {Herant}, \&
  {Colgate}}]{1999ApJ...516..892F}
{Fryer}, C., {Benz}, W., {Herant}, M., \& {Colgate}, S.~A. 1999, \apj, 516, 892

\bibitem[{{Fuller} \& {Ott}(2015)}]{2015MNRAS.450L..71F}
{Fuller}, J., \& {Ott}, C.~D. 2015, \mnras, 450, L71

\bibitem[{{Gossan} {et~al.}(2016){Gossan}, {Sutton}, {Stuver}, {Zanolin},
  {Gill}, \& {Ott}}]{2016PhRvD..93d2002G}
{Gossan}, S.~E., {Sutton}, P., {Stuver}, A., {et~al.} 2016, \prd, 93, 042002

\bibitem[{{Graham} {et~al.}(2018){Graham}, {Janish}, {Narayan}, {Rajendran}, \&
  {Riggins}}]{2018PhRvD..98k5027G}
{Graham}, P.~W., {Janish}, R., {Narayan}, V., {Rajendran}, S., \& {Riggins}, P.
  2018, \prd, 98, 115027

\bibitem[{{Graham} {et~al.}(2015){Graham}, {Rajendran}, \&
  {Varela}}]{2015PhRvD..92f3007G}
{Graham}, P.~W., {Rajendran}, S., \& {Varela}, J. 2015, \prd, 92, 063007

\bibitem[{{Hachisu}(1986)}]{1986ApJS...61..479H}
{Hachisu}, I. 1986, \apjs, 61, 479

\bibitem[{{Hansen} {et~al.}(2012){Hansen}, {Primas}, {Hartman}, {Kratz},
  {Wanajo}, {Leibundgut}, {Farouqi}, {Hallmann}, {Christlieb}, \&
  {Nilsson}}]{2012A&A...545A..31H}
{Hansen}, C.~J., {Primas}, F., {Hartman}, H., {et~al.} 2012, \aap, 545, A31

\bibitem[{{Hild} {et~al.}(2011){Hild}, {Abernathy}, {Acernese}, {Amaro-Seoane},
  {Andersson}, {Arun}, {Barone}, {Barr}, {Barsuglia}, {Beker}, {Beveridge},
  {Birindelli}, {Bose}, {Bosi}, {Braccini}, {Bradaschia}, {Bulik}, {Calloni},
  {Cella}, {Chassande Mottin}, {Chelkowski}, {Chincarini}, {Clark}, {Coccia},
  {Colacino}, {Colas}, {Cumming}, {Cunningham}, {Cuoco}, {Danilishin},
  {Danzmann}, {De Salvo}, {Dent}, {De Rosa}, {Di Fiore}, {Di Virgilio},
  {Doets}, {Fafone}, {Falferi}, {Flaminio}, {Franc}, {Frasconi}, {Freise},
  {Friedrich}, {Fulda}, {Gair}, {Gemme}, {Genin}, {Gennai}, {Giazotto},
  {Glampedakis}, {Gr{\"a}f}, {Granata}, {Grote}, {Guidi}, {Gurkovsky},
  {Hammond}, {Hannam}, {Harms}, {Heinert}, {Hendry}, {Heng}, {Hennes}, {Hough},
  {Husa}, {Huttner}, {Jones}, {Khalili}, {Kokeyama}, {Kokkotas}, {Krishnan},
  {Li}, {Lorenzini}, {L{\"u}ck}, {Majorana}, {Mandel}, {Mandic}, {Mantovani},
  {Martin}, {Michel}, {Minenkov}, {Morgado}, {Mosca}, {Mours},
  {M{\"u}ller─Ebhardt}, {Murray}, {Nawrodt}, {Nelson}, {Oshaughnessy}, {Ott},
  {Palomba}, {Paoli}, {Parguez}, {Pasqualetti}, {Passaquieti}, {Passuello},
  {Pinard}, {Plastino}, {Poggiani}, {Popolizio}, {Prato}, {Punturo}, {Puppo},
  {Rabeling}, {Rapagnani}, {Read}, {Regimbau}, {Rehbein}, {Reid}, {Ricci},
  {Richard}, {Rocchi}, {Rowan}, {R{\"u}diger}, {Santamar{\'\i}a}, {Sassolas},
  {Sathyaprakash}, {Schnabel}, {Schwarz}, {Seidel}, {Sintes}, {Somiya},
  {Speirits}, {Strain}, {Strigin}, {Sutton}, {Tarabrin}, {Th{\"u}ring}, {van
  den Brand}, {van Veggel}, {van den Broeck}, {Vecchio}, {Veitch}, {Vetrano},
  {Vicere}, {Vyatchanin}, {Willke}, {Woan}, \&
  {Yamamoto}}]{2011CQGra..28i4013H}
{Hild}, S., {Abernathy}, M., {Acernese}, F., {et~al.} 2011, CQGra, 28, 094013

\bibitem[{{Hillebrandt} {et~al.}(2013){Hillebrandt}, {Kromer}, {R{\"o}pke}, \&
  {Ruiter}}]{2013FrPhy...8..116H}
{Hillebrandt}, W., {Kromer}, M., {R{\"o}pke}, F.~K., \& {Ruiter}, A.~J. 2013,
  FrPhy, 8, 116

\bibitem[{{Janka}(2012)}]{2012ARNPS..62..407J}
{Janka}, H.-T. 2012, ARNPS, 62, 407

\bibitem[{{Jones} {et~al.}(2019){Jones}, {R{\"o}pke}, {Fryer}, {Ruiter},
  {Seitenzahl}, {Nittler}, {Ohlmann}, {Reifarth}, {Pignatari}, \&
  {Belczynski}}]{2019A&A...622A..74J}
{Jones}, S., {R{\"o}pke}, F.~K., {Fryer}, C., {et~al.} 2019, \aap, 622, A74

\bibitem[{{Kitaura} {et~al.}(2006){Kitaura}, {Janka}, \&
  {Hillebrandt}}]{2006A&A...450..345K}
{Kitaura}, F.~S., {Janka}, H.-T., \& {Hillebrandt}, W. 2006, \aap, 450, 345

\bibitem[{{Kouvaris} \& {Nielsen}(2015)}]{2015PhRvD..92f3526K}
{Kouvaris}, C., \& {Nielsen}, N.~G. 2015, \prd, 92, 063526

\bibitem[{{Lattimer} \& {Swesty}(1991)}]{1991NuPhA.535..331L}
{Lattimer}, J.~M., \& {Swesty}, F.~D. 1991, \nphysa, 535, 331

\bibitem[{{Lee} \& {Komatsu}(2010)}]{2010ApJ...718...60L}
{Lee}, J., \& {Komatsu}, E. 2010, \apj, 718, 60

\bibitem[{{Leung} {et~al.}(2011){Leung}, {Chu}, \& {Lin}}]{2011PhRvD..84j7301L}
{Leung}, S.-C., {Chu}, M.-C., \& {Lin}, L.-M. 2011, \prd, 84, 107301

\bibitem[{{Leung} {et~al.}(2015{\natexlab{a}}){Leung}, {Chu}, \&
  {Lin}}]{2015ApJ...812..110L}
---. 2015{\natexlab{a}}, \apj, 812, 110

\bibitem[{{Leung} {et~al.}(2015{\natexlab{b}}){Leung}, {Chu}, \&
  {Lin}}]{2015MNRAS.454.1238L}
---. 2015{\natexlab{b}}, \mnras, 454, 1238

\bibitem[{{Leung} {et~al.}(2013){Leung}, {Chu}, {Lin}, \&
  {Wong}}]{2013PhRvD..87l3506L}
{Leung}, S.-C., {Chu}, M.-C., {Lin}, L.-M., \& {Wong}, K.-W. 2013, \prd, 87,
  123506

\bibitem[{{Leung} \& {Nomoto}(2018)}]{2018ApJ...861..143L}
{Leung}, S.-C., \& {Nomoto}, K. 2018, \apj, 861, 143

\bibitem[{{Leung} \& {Nomoto}(2019)}]{2019arXiv190110007L}
---. 2019, arXiv e-prints, arXiv:1901.10007

\bibitem[{{Leung} {et~al.}(2019){Leung}, {Nomoto}, \&
  {Suzuki}}]{2019arXiv190111438L}
{Leung}, S.-C., {Nomoto}, K., \& {Suzuki}, T. 2019, arXiv e-prints,
  arXiv:1901.11438

\bibitem[{{Leung} {et~al.}(2019, in prep.){Leung}, {Zha}, {Chu}, {Lin}, \&
  {Nomoto}}]{2019LeungAIC}
{Leung}, S.~C., {Zha}, S., {Chu}, M.~C., {Lin}, L.~M., \& {Nomoto}, K. 2019, in
  prep.

\bibitem[{{Liebend{\"o}rfer}(2005)}]{2005ApJ...633.1042L}
{Liebend{\"o}rfer}, M. 2005, \apj, 633, 1042

\bibitem[{{Liu} {et~al.}(1994){Liu}, {Osher}, \& {Chan}}]{1994JCoPh.115..200L}
{Liu}, X.-D., {Osher}, S., \& {Chan}, T. 1994, JCoPh, 115, 200

\bibitem[{{Marek} {et~al.}(2006){Marek}, {Dimmelmeier}, {Janka}, {M{\"u}ller},
  \& {Buras}}]{2006A&A...445..273M}
{Marek}, A., {Dimmelmeier}, H., {Janka}, H.-T., {M{\"u}ller}, E., \& {Buras},
  R. 2006, \aap, 445, 273

\bibitem[{{Martinez} {et~al.}(2015){Martinez}, {Stovall}, {Freire}, {Deneva},
  {Jenet}, {McLaughlin}, {Bagchi}, {Bates}, \& {Ridolfi}}]{2015ApJ...812..143M}
{Martinez}, J.~G., {Stovall}, K., {Freire}, P.~C.~C., {et~al.} 2015, \apj, 812,
  143

\bibitem[{{Moenchmeyer} {et~al.}(1991){Moenchmeyer}, {Schaefer}, {Mueller}, \&
  {Kates}}]{1991A&A...246..417M}
{Moenchmeyer}, R., {Schaefer}, G., {Mueller}, E., \& {Kates}, R.~E. 1991, \aap,
  246, 417

\bibitem[{{Morozova} {et~al.}(2018){Morozova}, {Radice}, {Burrows}, \&
  {Vartanyan}}]{2018ApJ...861...10M}
{Morozova}, V., {Radice}, D., {Burrows}, A., \& {Vartanyan}, D. 2018, \apj,
  861, 10

\bibitem[{{Mukhopadhyay} \& {Schaffner-Bielich}(2016)}]{2016PhRvD..93h3009M}
{Mukhopadhyay}, P., \& {Schaffner-Bielich}, J. 2016, \prd, 93, 083009

\bibitem[{{M{\"u}ller} {et~al.}(2013){M{\"u}ller}, {Janka}, \&
  {Marek}}]{2013ApJ...766...43M}
{M{\"u}ller}, B., {Janka}, H.-T., \& {Marek}, A. 2013, \apj, 766, 43

\bibitem[{{Murphy} {et~al.}(2009){Murphy}, {Ott}, \&
  {Burrows}}]{2009ApJ...707.1173M}
{Murphy}, J.~W., {Ott}, C.~D., \& {Burrows}, A. 2009, \apj, 707, 1173

\bibitem[{{Nagakura} {et~al.}(2019){Nagakura}, {Furusawa}, {Togashi},
  {Richers}, {Sumiyoshi}, \& {Yamada}}]{2019ApJS..240...38N}
{Nagakura}, H., {Furusawa}, S., {Togashi}, H., {et~al.} 2019, \apjs, 240, 38

\bibitem[{{Nomoto} \& {Kondo}(1991)}]{1991ApJ...367L..19N}
{Nomoto}, K., \& {Kondo}, Y. 1991, \apjl, 367, L19

\bibitem[{{O'Connor}(2015)}]{2015ApJS..219...24O}
{O'Connor}, E. 2015, \apjs, 219, 24

\bibitem[{{Oertel} {et~al.}(2017){Oertel}, {Hempel}, {Kl{\"a}hn}, \&
  {Typel}}]{2017RvMP...89a5007O}
{Oertel}, M., {Hempel}, M., {Kl{\"a}hn}, T., \& {Typel}, S. 2017, RvMP, 89,
  015007

\bibitem[{{Ott}(2009)}]{2009CQGra..26f3001O}
{Ott}, C.~D. 2009, CQGra, 26, 063001

\bibitem[{{Ott} {et~al.}(2012){Ott}, {Abdikamalov}, {O'Connor}, {Reisswig},
  {Haas}, {Kalmus}, {Drasco}, {Burrows}, \& {Schnetter}}]{2012PhRvD..86b4026O}
{Ott}, C.~D., {Abdikamalov}, E., {O'Connor}, E., {et~al.} 2012, \prd, 86,
  024026

\bibitem[{{{\"O}zel} {et~al.}(2012){{\"O}zel}, {Psaltis}, {Narayan}, \& {Santos
  Villarreal}}]{2012ApJ...757...55O}
{{\"O}zel}, F., {Psaltis}, D., {Narayan}, R., \& {Santos Villarreal}, A. 2012,
  \apj, 757, 55

\bibitem[{{Pajkos} {et~al.}(2019){Pajkos}, {Couch}, {Pan}, \&
  {O'Connor}}]{2019arXiv190109055P}
{Pajkos}, M.~A., {Couch}, S.~M., {Pan}, K.-C., \& {O'Connor}, E.~P. 2019, arXiv
  e-prints, arXiv:1901.09055

\bibitem[{{Pan} {et~al.}(2018){Pan}, {Liebend{\"o}rfer}, {Couch}, \&
  {Thielemann}}]{2018ApJ...857...13P}
{Pan}, K.-C., {Liebend{\"o}rfer}, M., {Couch}, S.~M., \& {Thielemann}, F.-K.
  2018, \apj, 857, 13

\bibitem[{{Piro} \& {Kulkarni}(2013)}]{2013ApJ...762L..17P}
{Piro}, A.~L., \& {Kulkarni}, S.~R. 2013, \apjl, 762, L17

\bibitem[{{Piro} \& {Thompson}(2014)}]{2014ApJ...794...28P}
{Piro}, A.~L., \& {Thompson}, T.~A. 2014, \apj, 794, 28

\bibitem[{{Planck Collaboration} {et~al.}(2018){Planck Collaboration},
  {Aghanim}, {Akrami}, {Ashdown}, {Aumont}, {Baccigalupi}, {Ballardini},
  {Banday}, {Barreiro}, {Bartolo}, {Basak}, {Battye}, {Benabed}, {Bernard},
  {Bersanelli}, {Bielewicz}, {Bock}, {Bond}, {Borrill}, {Bouchet}, {Boulanger},
  {Bucher}, {Burigana}, {Butler}, {Calabrese}, {Cardoso}, {Carron},
  {Challinor}, {Chiang}, {Chluba}, {Colombo}, {Combet}, {Contreras}, {Crill},
  {Cuttaia}, {de Bernardis}, {de Zotti}, {Delabrouille}, {Delouis}, {Di
  Valentino}, {Diego}, {Dor{\'e}}, {Douspis}, {Ducout}, {Dupac}, {Dusini},
  {Efstathiou}, {Elsner}, {En{\ss}lin}, {Eriksen}, {Fantaye}, {Farhang},
  {Fergusson}, {Fernandez-Cobos}, {Finelli}, {Forastieri}, {Frailis},
  {Franceschi}, {Frolov}, {Galeotta}, {Galli}, {Ganga}, {G{\'e}nova-Santos},
  {Gerbino}, {Ghosh}, {Gonz{\'a}lez-Nuevo}, {G{\'o}rski}, {Gratton},
  {Gruppuso}, {Gudmundsson}, {Hamann}, {Hand ley}, {Herranz}, {Hivon}, {Huang},
  {Jaffe}, {Jones}, {Karakci}, {Keih{\"a}nen}, {Keskitalo}, {Kiiveri}, {Kim},
  {Kisner}, {Knox}, {Krachmalnicoff}, {Kunz}, {Kurki-Suonio}, {Lagache},
  {Lamarre}, {Lasenby}, {Lattanzi}, {Lawrence}, {Le Jeune}, {Lemos},
  {Lesgourgues}, {Levrier}, {Lewis}, {Liguori}, {Lilje}, {Lilley}, {Lindholm},
  {L{\'o}pez-Caniego}, {Lubin}, {Ma}, {Mac{\'\i}as-P{\'e}rez}, {Maggio},
  {Maino}, {Mandolesi}, {Mangilli}, {Marcos-Caballero}, {Maris}, {Martin},
  {Martinelli}, {Mart{\'\i}nez-Gonz{\'a}lez}, {Matarrese}, {Mauri}, {McEwen},
  {Meinhold}, {Melchiorri}, {Mennella}, {Migliaccio}, {Millea}, {Mitra},
  {Miville-Desch{\^e}nes}, {Molinari}, {Montier}, {Morgante}, {Moss}, {Natoli},
  {N{\o}rgaard-Nielsen}, {Pagano}, {Paoletti}, {Partridge}, {Patanchon},
  {Peiris}, {Perrotta}, {Pettorino}, {Piacentini}, {Polastri}, {Polenta},
  {Puget}, {Rachen}, {Reinecke}, {Remazeilles}, {Renzi}, {Rocha}, {Rosset},
  {Roudier}, {Rubi{\~n}o-Mart{\'\i}n}, {Ruiz-Granados}, {Salvati}, {Sandri},
  {Savelainen}, {Scott}, {Shellard}, {Sirignano}, {Sirri}, {Spencer},
  {Sunyaev}, {Suur-Uski}, {Tauber}, {Tavagnacco}, {Tenti}, {Toffolatti},
  {Tomasi}, {Trombetti}, {Valenziano}, {Valiviita}, {Van Tent}, {Vibert},
  {Vielva}, {Villa}, {Vittorio}, {Wand elt}, {Wehus}, {White}, {White},
  {Zacchei}, \& {Zonca}}]{2018arXiv180706209P}
{Planck Collaboration}, {Aghanim}, N., {Akrami}, Y., {et~al.} 2018, arXiv
  e-prints, arXiv:1807.06209

\bibitem[{{Radice} {et~al.}(2019){Radice}, {Morozova}, {Burrows}, {Vartanyan},
  \& {Nagakura}}]{2019ApJ...876L...9R}
{Radice}, D., {Morozova}, V., {Burrows}, A., {Vartanyan}, D., \& {Nagakura}, H.
  2019, \apj, 876, L9

\bibitem[{{Reisswig} \& {Pollney}(2011)}]{2011CQGra..28s5015R}
{Reisswig}, C., \& {Pollney}, D. 2011, Classical and Quantum Gravity, 28,
  195015

\bibitem[{{Ren} {et~al.}(2018){Ren}, {Zhao}, {Abdukerim}, {Chen}, {Chen},
  {Cui}, {Fang}, {Fu}, {Giboni}, {Giuliani}, {Gu}, {Guo}, {Han}, {He}, {Huang},
  {He}, {Huang}, {Huang}, {Ji}, {Ju}, {Li}, {Lin}, {Liu}, {Liu}, {Ma}, {Mao},
  {Ni}, {Ning}, {Tan}, {Wang}, {Wang}, {Wang}, {Wang}, {Wang}, {Wu}, {Xia},
  {Xiao}, {Xie}, {Yan}, {Yang}, {Yang}, {Yu}, {Yue}, {Zhang}, {Zhou}, {Zhou},
  {Zheng}, {Zhou}, \& {PandaX-II Collaboration}}]{2018PhRvL.121b1304R}
{Ren}, X., {Zhao}, L., {Abdukerim}, A., {et~al.} 2018, \prl, 121, 021304

\bibitem[{{Richers} {et~al.}(2017){Richers}, {Ott}, {Abdikamalov}, {O'Connor},
  \& {Sullivan}}]{2017PhRvD..95f3019R}
{Richers}, S., {Ott}, C.~D., {Abdikamalov}, E., {O'Connor}, E., \& {Sullivan},
  C. 2017, \prd, 95, 063019

\bibitem[{{Ruiter} {et~al.}(2019){Ruiter}, {Ferrario}, {Belczynski},
  {Seitenzahl}, {Crocker}, \& {Karakas}}]{2019MNRAS.484..698R}
{Ruiter}, A.~J., {Ferrario}, L., {Belczynski}, K., {et~al.} 2019, \mnras, 484,
  698

\bibitem[{{Schwab} {et~al.}(2017){Schwab}, {Bildsten}, \&
  {Quataert}}]{2017MNRAS.472.3390S}
{Schwab}, J., {Bildsten}, L., \& {Quataert}, E. 2017, \mnras, 472, 3390

\bibitem[{{Schwab} {et~al.}(2010){Schwab}, {Podsiadlowski}, \&
  {Rappaport}}]{2010ApJ...719..722S}
{Schwab}, J., {Podsiadlowski}, P., \& {Rappaport}, S. 2010, \apj, 719, 722

\bibitem[{{Schwab} \& {Rocha}(2019)}]{2019ApJ...872..131S}
{Schwab}, J., \& {Rocha}, K.~A. 2019, \apj, 872, 131

\bibitem[{{Shen} {et~al.}(2011){Shen}, {Toki}, {Oyamatsu}, \&
  {Sumiyoshi}}]{2011ApJS..197...20S}
{Shen}, H., {Toki}, H., {Oyamatsu}, K., \& {Sumiyoshi}, K. 2011, \apjs, 197, 20

\bibitem[{{Skinner} {et~al.}(2016){Skinner}, {Burrows}, \&
  {Dolence}}]{2016ApJ...831...81S}
{Skinner}, M.~A., {Burrows}, A., \& {Dolence}, J.~C. 2016, \apj, 831, 81

\bibitem[{{Steiner} {et~al.}(2013){Steiner}, {Hempel}, \&
  {Fischer}}]{2013ApJ...774...17S}
{Steiner}, A.~W., {Hempel}, M., \& {Fischer}, T. 2013, \apj, 774, 17

\bibitem[{{Torres-Forn{\'e}} {et~al.}(2019){Torres-Forn{\'e}},
  {Cerd{\'a}-Dur{\'a}n}, {Obergaulinger}, {M{\"u}ller}, \&
  {Font}}]{2019arXiv190210048T}
{Torres-Forn{\'e}}, A., {Cerd{\'a}-Dur{\'a}n}, P., {Obergaulinger}, M.,
  {M{\"u}ller}, B., \& {Font}, J.~A. 2019, arXiv e-prints, arXiv:1902.10048

\bibitem[{{Vartanyan} {et~al.}(2019){Vartanyan}, {Burrows}, {Radice},
  {Skinner}, \& {Dolence}}]{2019MNRAS.482..351V}
{Vartanyan}, D., {Burrows}, A., {Radice}, D., {Skinner}, M.~A., \& {Dolence},
  J. 2019, \mnras, 482, 351

\bibitem[{{Wang}(2018)}]{2018MNRAS.481..439W}
{Wang}, B. 2018, \mnras, 481, 439

\bibitem[{{Xiang} {et~al.}(2014){Xiang}, {Jiang}, {Zhang}, \&
  {Yang}}]{2014PhRvC..89b5803X}
{Xiang}, Q.-F., {Jiang}, W.-Z., {Zhang}, D.-R., \& {Yang}, R.-Y. 2014, \prc,
  89, 025803

\bibitem[{{Yoon} \& {Langer}(2004)}]{2004A&A...419..623Y}
{Yoon}, S.-C., \& {Langer}, N. 2004, \aap, 419, 623

\end{thebibliography}
	
\end{document}